\documentclass[12pt]{article}
\usepackage{eurosym}
\usepackage{amsfonts}
\usepackage{amsmath}
\usepackage{amscd}
\usepackage{amssymb}
\usepackage{graphicx}
\usepackage{wasysym}
\usepackage{pifont}
\usepackage[hyphens]{url}
\usepackage[colorlinks=true,urlcolor=black]{hyperref}
\usepackage{hyperref}
\usepackage{boxedminipage}
\usepackage{color}
\usepackage{multirow}
\usepackage[all]{xy}
\usepackage{ulem}
\usepackage{cancel}
\usepackage{orcidlink}

\newcommand{\beqa}{\begin{eqnarray}}
\newcommand{\eeqa}{\end{eqnarray}}

\hypersetup{
backref = true,
pagebackref = true,
hyperindex = true,
colorlinks = true,
breaklinks = true,
urlcolor = blue,
linkcolor = blue,
bookmarks = true,
bookmarksopen = true,
citecolor=red,
}
\numberwithin{equation}{section}

\begin{document}

\title{ Kac-Moody Algebras on Soft Group Manifolds}
\author{}
\maketitle

\begin{center}
{\large Rutwig Campoamor-Stursberg$^{1 \ast \;\orcidlink{0000-0003-2907-8533}%
}$, Alessio Marrani $^{2,3,4 \dagger\; \orcidlink{0000-0002-7597-1050}}$ ,}
\\[0pt]
and {\large Michel Rausch de Traubenberg$^{5 \ddagger \; %
\orcidlink{0000-0001-5045-2353}}$,}\\[5pt]
\bigskip \bigskip

$^1$ Instituto de Matem\'atica Interdisciplinar and Dp.to de Geometr\'\i a y
Topolog\'\i a, \\[0pt]
UCM, E-28040 Madrid, Spain\\[5pt]

$^2$ Elementar, Divisione Ricerca e Sviluppo, Galleria Enzo Tortora 21,
I-10121 Torino, Italy \\[5pt]

$^3$ Dipartimento di Management `Valter Cantino', Universit\'{a} di Torino,
\\[0pt]
Corso Unione Sovietica 218bis, I-10134 Torino, Italy\\[5pt]

$^4$ Department of Physics, Astronomy and Mathematics, University of
Hertfordshire, \\[0pt]
Hatfield, Hertfordshire, AL10 9AB, UK\\[5pt]

$^5$ Universit\'e de Strasbourg, CNRS, IPHC UMR7178, F-67037 Strasbourg
Cedex, France\\[5pt]
\end{center}

\bigskip

\bigskip

\begin{abstract}
Within the so-called group geometric approach to (super)gravity and
(super)string theories, any compact Lie group manifold $G_{c}$ can be
smoothly deformed into a group manifold $G_{c}^{\mu }$ (locally
diffeomorphic to $G_{c}$ itself), which is `soft', namely, based on a
non-left-invariant, intrinsic one-form Vielbein $\mu $, which violates the
Maurer-Cartan equations and consequently has a non-vanishing associated
curvature two-form. Within the framework based on the above deformation
(`softening'), we show how to construct an infinite-dimensional
(infinite-rank), generalized Kac-Moody (KM) algebra associated to $%
G_{c}^{\mu }$, starting from the generalized KM algebras associated to $%
G_{c} $. As an application, we consider KM algebras associated to deformed
manifolds such as the `soft' circle, the `soft' two-sphere and the `soft'
three-sphere. While the generalized KM algebra associated to the deformed
circle is trivially isomorphic to its undeformed analogue, and hence not
new, the `softening' of the two- and three- sphere includes squashed
manifolds (and in particular, the so-called Berger three-sphere) and yields
to non-trivial results.

\bigskip

\bigskip

\bigskip

\bigskip

\bigskip

\noindent $^\ast$ rutwig@ucm.es

\noindent $^\dagger$ jazzphyzz@gmail.com

\noindent $^\ddagger$ Michel.Rausch@iphc.cnrs.fr
\end{abstract}

\newpage

\tableofcontents

\newpage

%
%
%
%
%
%
%
%
%

%
%
%
%
%
%
%
%

\section{Introduction}

Kac-Moody (KM) algebras, a fascinating generalization of finite-dimensional
Lie algebras, have been, and still are, a cornerstone of modern Theoretical
Physics and Mathematical Physics. Since the early 80's, their emergence in
various domains, from string theory and conformal field theory (CFT) to
Yang-Mills theory, integrable systems and quantum groups, underscores their
fundamental role in understanding the symmetry and structure of physical
systems (see for instance \cite{DiFrancesco:1997nk}).

The affine extension of the loop algebra of smooth maps from the unit circle
$\mathbb{S}^{1}$ into a simple Lie group allows to construct KM (or, better,
affine) Lie algebras \cite{Kac:1967jr}\nocite%
{Kac:1990gs,Moody:1966gf,Pressley:1988qk}-\cite{Goddard:1986bp}, as an
alternative to their axiomatic construction (which is instead essentially
based on the relaxation of the property of positive definiteness of the
corresponding Cartan matrix). Generally, KM algebras are classified into
three types: I) \textit{finite type}, corresponding to finite-dimensional
semisimple Lie algebras; II) \textit{affine type}, related to loop algebras
and key in two-dimensional CFT; III) \textit{indefinite type}, the least
understood class, potentially linked to hyperbolic symmetries. This latter
type plays an important role in cosmology, as for instance in the so-called
Belinskii-Khalatnikov-Lifshitz (BKL) scenario, in which the chaotic
oscillations of the metric of space-time near a singularity resemble a
billiard motion governed by hyperbolic KM algebras (\cite{Belinsky:1970ew};
for applications in supergravity, see e.g. \cite{Fre:2005gvm} and Refs.
therein).

In two-dimensional CFT's, affine KM algebras describe the local symmetry.
These algebras provide a rich framework for studying primary fields and
their correlation functions, and also discriminate among the various
universality classes in the statistical physics of critical phenomena. On
the other hand, in string theory the constraints of the string's world-sheet
are governed by the Virasoro algebra, which is a central extension of the
Witt algebra (i.e., of the algebra of polynomial vector fields on $\mathbb{S}%
^{1}$), and is intimately connected to affine Kac-Moody algebras via the
Sugawara construction (see e.g. \cite{Belavin:1984vu,Fuks}).

Along the years, a number of extensions and generalizations of KM algebras
has been introduced and applied in a variety of contexts. Just to name a
few, it is here worth mentioning quasi-simple Lie algebras \cite%
{HOEGHKROHN1990106}, the generalization of KM algebras introduced in \cite%
{Frappat:1989gn}, and the extended Borcherds KM algebras \cite%
{BORCHERDS1991330}: these latter allow for imaginary simple roots, and find
applications in supergravity (see e.g. \cite{Henneaux:2010ys}) as well as
within the Monstrous Moonshine \cite{monstrous}\nocite{Borcherds:1992jjg}-%
\cite{Gannon:2007mfj}.

The Witt and the Virasoro algebras, and more generally the KM algebras, are
intimately related to the compact one-dimensional manifold $\mathbb{S}^{1}$.
This fact hints for the following tantalizing question: \textit{do other
more general, infinite-dimensional KM algebras, related to
higher-dimensional compact manifolds, exist?} This is ultimately motivated
also by higher-dimensional physical theories, in which harmonic expansions
\textit{\`{a} la Kaluza-Klein} play a crucial role (see e.g. \cite%
{Salam:1981xd}\nocite{Bailin:1987jd}-\cite{KK3}, with the latter reference
being motivated in the supergravity context).

The answer to the above question is positive, and was formulated in \cite%
{mrm} (see also \cite{rmm2}-\nocite{Campoamor-Stursberg:2022ane,Campoamor-Stursberg:2022lyx}\cite{rm}), in which
a broad class of `generalized KM algebras', based on spaces of
differentiable maps from compact manifolds $\mathcal{M}$ to compact Lie
groups $G$, was introduced and investigated, then restricting $\mathcal{M}$
to be a compact Lie group manifold itself ($\mathcal{M}=G_{c}$) or a coset
thereof ($\mathcal{M}=G_{c}/H$, where $H$ is a closed subgroup of $G_{c}$);
such a restriction implied major simplifications in the treatment, because
the harmonic functions on $\mathcal{M}$ could then be classified in terms of
the representation theory of $G_{c}$ itself, and subsequently the Peter-Weyl
theorem could be applied. In the past, such generalized KM
algebras have been considered by various authors for specific manifolds,
such as the two-sphere \cite{Bars:1983uv} or the $n$-tori \cite{HOEGHKROHN1990106,Bars:1983uv,MRT,Harada:2020exi}; it is also
here worth mentioning that, within the formulation of an extension of
general relativity to closed string field theory, in \cite{ca1} the
possibility of generalizing KM algebras by replacing $\mathbb{S}^{1}$ with a
compact coset $G_{c}/H$ was put forward.

All the aforementioned generalization of KM algebras lie outside Kac's
classification of KM algebras \cite{Kac:1990gs}; nevertheless, they can
ultimately be regarded as generalizations of affine Lie algebras, admitting
roots but not simple roots (and thus, in many cases, they do not admit a
Cartan matrix at all). Moreover, unlike the usual KM algebras (whose
generators are only iteratively known, level by level, by means of the
Chevalley-Serre relations), the generators of such generalized KM algebras
can all be constructed explicitly.

The further, somewhat natural, generalization of such algebras consisting in
the relaxation of the condition of compactness of $\mathcal{M}$, has been
considered in \cite{ram} (see also \cite{Campoamor-Strusberg:2024kpl}), by
focusing on the toy model provided by $\mathcal{M}=SL(2,\mathbb{R})$ and on
the related symmetric coset $SL(2,\mathbb{R})/U(1)$. The lack of compactness
makes the harmonic analysis on such manifolds highly non-trivial, and one
must resort to different methods (with respect to the ones exploited in \cite%
{mrm}) in order to extract suitable bases of the corresponding Hilbert
spaces; in particular, the treatment given in \cite{ram} has a twofold
nature: on the one hand, the Peter-Weyl theorem had to be superseded by the
Plancherel theorem (generally displaying discrete and continuous series of
representations of $SL(2,\mathbb{R})$ itself), while on the other hand a
Hilbert basis on the space of square-integrable functions $L^{2}(SL(2,%
\mathbb{R}))$ was identified. The appearance of $SL(2,\mathbb{R})/U(1)$ as
the target space of one complex scalar field in the bosonic sector of some
Maxwell-Einstein supergravity theories in $D=3+1$ space-time dimensions
would hint at an application of the resulting generalized KM algebra $%
\widehat{\mathfrak{g}}(SL(2,\mathbb{R})/U(1))$ in the context of
supergravity (as preliminarily discussed in \cite{ram}).\medskip

In this paper, we present a further generalization of the above class of
infinite-dimensional (infinite-rank) generalized KM algebras, considering
the compact group manifold $G_{c}$ to be `deformed' into a so-called `soft'
group manifold $G_{c}^{\mu }$, locally diffeomorphic to $G_{c}$ itself (the
same can be done for cosets of $G_{c}$, as well). Usually, within the
so-called group-geometric approach to (super)gravity and superstring
theories, group manifolds are `softened' in order to become domains of
definition of gravitational dynamical fields, in such a way to regard $%
G_{c}^{\mu }$ as a vacuum configuration of a gravitational theory. Group
geometry provides a natural and unified formulation of gravity and gauge
theories, such that the invariances of both are interpreted as
diffeomorphisms on a suitable group manifold. This geometrical framework
provides a systematic algorithm for the gauging of Lie algebras and the
construction of (super)gravity and (super)strings Lagrangians, and was
extensively developed by the research group led by Tullio Regge, and later
by his disciples Riccardo D'Auria and Pietro Fr\'{e}, in Turin, starting
more than fifty years ago; some sketchy presentation of the main facts will
be given in Sec. 2.2, but for more details we address the reader to the
lectures \cite{cas}, as well as to the the first of the three books on
supergravity and string theories written by Castellani, D'Auria and Fr\'{e}
\cite{Castellani:1991et} (see in particular Sec. I.3 therein). The
`softening' essentially amounts to deforming the original, rigid structure
of the group manifold $G_{c}$, whose left- or right- invariant vector fields
and one-forms have (in a given chart) a fixed coordinate dependence, and
whose Riemannian geometry is (locally) fixed in terms of the structure
constants of (the Lie algebra $\mathfrak{g}_{c}$ of) $G_{c}$ itself.

In other words, in this paper we consider the compact group manifold $G_{c}$
be deformed (`softened') into the `soft' compact group manifold $G_{c}^{\mu
} $, which is then potentially able to describe non-trivial physical
configurations. This should be regarded as a crucial step towards the
application of the generalized KM algebras under consideration in
(super)gravity and (super)string theories. As explicit examples, we will
consider the deformations (`softenings') of the circle (one-sphere) $\mathbb{%
S}^{1}=SO(2)\simeq U(1)$, of the two-sphere $\mathbb{S}^{2}=SO(3)/SO(2)%
\simeq SU(2)/U(1)=\mathbb{C}P^{1}$, and of the three-sphere $\mathbb{S}%
^{3}=SO(4)/SO(3)\simeq \left( SU(2)\times SU(2)\right) /SU(2)\simeq SU(2)$,
respectively into the deformed circle $\mathbb{S}_{F}^{1}$, deformed
two-sphere $\mathbb{S}_{F}^{2}$ and deformed three-sphere $\mathbb{S}%
_{F}^{3} $, all enjoying remarkable physical applications. As we will see,
while the deformation of $\mathbb{S}^{1}$ is actually immaterial, the
`softened' manifolds $\mathbb{S}_{F}^{2}$ and $\mathbb{S}_{F}^{3}$ include,
as special cases, the so-called \textit{squashed} two-sphere $\widetilde{%
\mathbb{S}}^{2} $ and \textit{squashed} three-sphere (also named \textit{%
Berger} three-sphere) $\widetilde{\mathbb{S}}^{3}$.\bigskip

The plan of the paper is as follows: Sec. 2 contains the general results
described above, and it is split into three subsections. In Sec. 2.1, we
recall the construction of a generalized KM algebra associated to a compact
group manifold $G_{c}$ (or to a coset thereof). Then, in Sec. 2.2, we recall
the basic facts of the deformation (`softening') procedure yielding from $%
G_{c}$ to the `soft' group manifold $G_{c}^{\mu }$. Subsequently, in Sec.
2.3 we associate a generalized KM algebra to $G_{c}^{\mu }$, starting from
the generalized KM algebras associated to $G_{c}$. Sec. 3 presents the most
elementary example of such a procedure, pertaining to the circle $\mathbb{S}%
^{1}$, in which case the generalized KM algebra associated to the deformed
circle $\mathbb{S}_{F}^{1}$ is trivially isomorphic to its undeformed
analogue; this is not surprising, since the topological classification of
one-dimensional closed curves shows that \textit{all} such curves are
topologically equivalent to $\mathbb{S}^{1} $. Then, Sec. 4 presents a
detailed treatment of the three-sphere $\mathbb{S}^{3}$, whose `softened'
version includes the so-called squashed three-sphere $\widetilde{\mathbb{S}}%
^{3}$. Finally, by removing the dependence on one angle, Sec. 5 deals with
the two-sphere $\mathbb{S}^{2}$ and its `softening' $\mathbb{S}_{F}^{2}$,
which includes the so-called squashed two-sphere $\widetilde{\mathbb{S}}^{2}$%
. Some conclusive remarks are made in Sec. 6, and the paper is concluded by
an Appendix, which recalls some basic facts on the Maurer-Cartan one-forms
of $SU(2)$.

\section{Kac-Moody (KM) algebras on `soft' Lie group manifolds}

Let $G_{c}$ be a compact Lie group. In this section we recall the salient
steps for the construction of a Kac-Moody (KM) algebra associated to $G_{c}$
(for more details see \cite{mrm}). In a second part, we turn to the
construction of a KM algebra on a deformation (more specifically, a
`softening', see below) of $G_{c}$.

\subsection{A KM algebra on $G_c$}

Let $\mathfrak{g}=\{T_1,\cdots, T_d\}$ be a $d-$dimensional simple Lie
algebra (complex or real) with Lie brackets
\begin{equation}
\big[T_a,T_b \big]=i f_{ab}{}^c T_c, \label{eq:g}
\end{equation}
and Killing form
\begin{equation}
\big<T_a, T_b\big>_0=\eta_{ab}=\text{tr}(\text{ad}(T_a) \text{ad}(T_b))\ .
\notag
\end{equation}

Further, let $G_{c}$ be a compact Lie group. Hence, $G_{c}$ is a
compact group manifold, that we assume to be of dimension $n$. Let $%
m^{A}=(\varphi ^{i},\theta ^{r})$ with $i=1,\cdots ,p;r=1,\cdots ,q$ and $%
n=p+q$ be a parameterisation of $G_{c}$; this split of $%
m^{A}$ is such that the matrix elements are periodic in the $\varphi ^{i}$'s
and not periodic in the $\theta ^{r}$'s\footnote{%
For instance, the spherical coordinates' parametrization of the (unit) $%
\mathbb{S}^{2}$ has $\varphi \in \lbrack 0,2\pi )$ and $\cos \theta \in
\lbrack -1,1].$}. Within this parameterisation, a generic element
(connected to the identity) of $G_{c}$ is given by
\begin{equation*}
g(m)=e^{im^{A}J_{A}}\ ,
\end{equation*}%
where $J_{A},A=1,\cdots ,n$ are the generators of $\mathfrak{g}_{c}$, the
Lie algebra of $G_{c}$, not to be confused with the generators of $\mathfrak{%
g}$. Then, denote the coordinates of a group element (in a local coordinate
chart) by
\begin{equation*}
g(m)^{M}\equiv m^{M}\ ,\ \ M=1,\cdots ,n.
\end{equation*}%
The indices $A,B,\cdots $ are tangent space indices, \textit{i.e.}, flat
indices, whilst the indices $M,N,\cdots $ are world indices, \textit{i.e.},
curved indices. The Vierbein one-form associated to the aforementioned parameterisation reduces to
\begin{equation*}
e(m)=g(m)^{-1}\;\mathrm{d}g(m)\ ,
\end{equation*}%
and satisfies the Maurer-Cartan equation
\begin{equation}
\text{d}e+e\wedge e=0\ . \label{eq:e}
\end{equation}%
The components of Vielbein are
\begin{equation*}
e^{A}(m)=e_{M}{}^{A}(m)\;\text{d}m^{M}\ .
\end{equation*}%
Therefore the metric tensor on $G_{c}$ is
\begin{equation}
g_{MN}=e_{M}{}^{A}(m)e_{N}{}^{B}(m)\delta _{AB}\ . \label{eq:gG}
\end{equation}

We consider now the Hilbert space $L^{2}(G_{c})$, \textit{i.e.}, the set of
square integrable functions on $G_{c}$, endowed with the scalar product:
\begin{equation}
(f,g)=\frac{1}{V}\int_{G_{c}}\sqrt{g}\;\text{d}\varphi ^{p}\text{d}\theta
^{q}\overline{f(\varphi ,\theta )}g(\varphi ,\theta ), \label{eq:PSGc}
\end{equation}%
where $g=\det (g_{MN})$ and $V$ is the volume of $G_{c}$. Since $G_{c}$ is
compact, its unitary representations are finite dimensional and it turns out
that, once correctly normalized, the set of all matrix elements constitutes
an orthonormal Hilbert basis of $L^{2}(G_{c})$. This is the Peter--Weyl
Theorem \cite{PW}. Introduce $\mathcal{I}$ the minimal countable set of
labels required to identify the states unambiguously. With this notation,
one can write the matrix elements as $\rho _{I}(\varphi ,\theta )$ (\cite%
{mrm}, see also below) and the Hilbert basis of $L^{2}(G_{c})$ is given by
\begin{equation}
\mathcal{B}=\Big\{\rho _{I}(\varphi ,\theta )\ ,\ \ I\in \mathcal{I}\Big\}\ , \label{eq:B}
\end{equation}%
and the matrix elements are orthonormal with respect to the scalar product %
\eqref{eq:PSGc}
\begin{equation*}
(\rho _{I},\rho _{J})=\delta _{IJ}\ .
\end{equation*}%
Since the product of two different elements of the basis $\mathcal{B}$ is
square integrable, we have the following decomposition
\begin{equation}
\rho _{I}(\varphi ,\theta )\rho _{J}(\varphi ,\theta )=c_{IJ}{}^{K}\rho
_{K}(\varphi ,\theta ) \label{eq:rhorho}
\end{equation}%
where the coefficients $c_{IJ}{}^{K}=c_{JI}{}^{K}$ are the Clebsch-Gordan
coefficients of $G_{c}$ \cite{mrm}.

The parameterisation $m^A=(\varphi^i,\theta^r)$ leads naturally to a
differential realization of the Lie algebra $\mathfrak{g}_c$ for the
generators of the left and right action. Let $L_A$ (resp. $R_A$) $A=1,\cdots
n$, be the generators of the left (resp. right) action satisfying
\begin{eqnarray}
\big[L_A,L_B\big]=i C_{AB}{}^C L_C \ , \ \ \big[R_A,R_B\big]=i C_{AB}{}^C
R_C \ , \ \ \big[L_A,R_B\big]=0 \ ,  \notag
\end{eqnarray}
where $C_{AB}{}^C$ are the structure constants of $\mathfrak{g}_c$. Recall
how one can construct in an explicit way the generators $L_A$ and $R_A$. Let
$\mathcal{M}$ be a Riemannian manifold. Let $m^M, M=1,\cdots, \dim \mathcal{M%
}$ be a parameterisation of $\mathcal{M}$. The manifold $\mathcal{M}$ is
endowed with a metric $g$ and let $\nabla_M$ be the corresponding covariant
derivative \cite{Gol}. Consider a coordinate transformation $m^M \to m^M +
\epsilon \xi^M$ with $\epsilon \sim 0$. This transformation is an isometry,
\textit{i.e.}, it preserves the metric if we have the Killing equation:
\begin{eqnarray}  \label{eq:Killing}
\nabla_M \xi_N + \nabla_N \xi_M=0 \ ,
\end{eqnarray}
where $\xi_M = g_{MN}\xi^N$. For any solution of the Killing equation $%
\xi^M_A(m)$, the differential operator
\begin{eqnarray}
J_A= -i\xi_A^M(m)\; \frac{\partial}{\partial m^M}  \notag
\end{eqnarray}
leaves the metric invariant. When the manifold corresponds to a group
manifold, the generators $L_A, R_A$ (corresponding of the left and right
action of $G_c$ onto itself) are solutions of the Killing equation %
\eqref{eq:Killing}.

Furthermore, recall that if $\omega$ is a one-form under the isometry $J_A$,
we have
\begin{eqnarray}
\delta_A \omega_M=-\Big(\xi_A^N(m)\frac{\partial\omega_M}{\partial m^N}
+\omega_N \frac{\partial\xi_A^N(m)}{\partial m^M}\Big)\ .  \notag
\end{eqnarray}
Thus $\omega$ is invariant under the action of $J_A$ if we have
\begin{eqnarray}  \label{eq:inv1form}
\xi_A^N(m)\frac{\partial\omega_M}{\partial m^N} +\omega_N \frac{%
\partial\xi_A^N(m)}{\partial m^M}=0 \ .
\end{eqnarray}

Let $I=(L,Q,R)$ with $L=(\ell_1,\cdots,\ell_{\frac12(n-r_c)}),R=(r_1,%
\cdots,r_{\frac12(n-r_c)})$ and $Q=(q_q,\cdots,q_{r_c})$, where $r_c$ is the
rank of $\mathfrak{g}_c$. Indeed, a unitary representation is specified by
the $r_c$ eigenvalues of the primitive Casimir operators, and any vector of
a given representation is unambiguously specified by the eigenvalues of $%
\frac12(n-r_c)$ commuting Hermitean operators \cite{mrm}. Thus $L,Q,R$
constitute the minimal set of indices to specify any vector in the basis $%
\mathcal{B}$ \eqref{eq:B} (see \cite{Ra,Sh3,C97} for details). The action
of $L_A,R_A$ reads
\begin{eqnarray}
L_A \rho_{LQR}(\varphi,\theta)&=& C^Q_{AL}{}^{L^{\prime }}\rho_{L^{\prime
}QR}(\varphi,\theta)  \notag \\
R_A \rho_{LQR}(\varphi,\theta)&=& C^Q_{AR}{}^{R^{\prime }}\rho_{LQR^{\prime
}}(\varphi,\theta) \   \notag
\end{eqnarray}
where $C^Q_{AL}{}^{L^{\prime }}$ (resp. $C^Q_{AR}{}^{R^{\prime }}$) are the
matrix elements of the left (right) action for the representation specified
by $Q$.

We next define a space of smooth mappings from $G_c$ into $\mathfrak{g}$ as
\begin{eqnarray}  \label{eq:loop}
\mathfrak{g}(G_c) =\Big\{T_{aI} = T_a \rho_I (\varphi, \theta) \ ,\ \ a =
1,\cdots, d \ , \ \ I \in \mathcal{I} \Big\} \ ,
\end{eqnarray}
which inherits the structure of a Lie algebra
\begin{eqnarray}  \label{eq:hatg}
\big[T_{aI} , T_{bJ}\big] = i f_{ab}{}^c c_{IJ}{}^K T_{cK} .
\end{eqnarray}
This, in particular means that the Lie brackets of $\mathfrak{g}(G_c)$ are
obtained in terms of the structure constants $f_{ab}{}^c$ of $\mathfrak{g}$ and the
Clebsch-Gordan coefficients $c_{IJ}{}^K$ of $G_c$.

The last step in the construction of the KM algebra associated to $G_c$ is
to introduce Hermitean operators and central charges in duality. Of course,
as the variables $\varphi^i$ are periodic, the operators $-i
\partial_{\varphi^ i}$ are Hermitean. However, additional Hermitian
operators can be considered. For instance, the $2 r_c$ set of commuting
operators obtained from $L_A,R_B$ and corresponding to the Cartan subalgebra
of $\mathfrak{g}_c$ constitutes a set of $2 n_c$ Hermitean commuting
operators. In some case, however, this number can be larger. The existence
of these additional commuting operators has been analysed in \cite{mrm}.
Denote $D_i, i=2 r_c\le 1,\cdots, r \le n$ the set of commuting Hermitean
operators ($r=n$ only in the case of the $n-$tori $\mathbb{T}_n=U(1)^n$).
The elements $T_{aI}$ are eigenfunctions of $D_i$:
\begin{eqnarray}  \label{eq:DT}
\big[D_i, T_{aI}\big]&=& I(i) T_{aI} \ .
\end{eqnarray}

There exists in principle an infinite number of central charges, hence we
limit ourselves to $r$ central charges $k_1,\cdots,k_r$ in duality with the
Hermitean operators, and given by the two-cocycle (see also \cite{ps,bt}):
\begin{eqnarray}  \label{eq:k}
\omega_i(T_{aI},T_{bJ})&=& \frac{k_i}V \int_{G_c} \sqrt{g} \text{d}^p
\varphi \text{d}^ q \theta \Big<T_{aI}, D_i T_{bJ}\Big>_0  \notag \\
&=&k_i J(i)\eta_{ab} \eta_{IJ}
\end{eqnarray}
where $\eta_{IJ}= \pm \delta_{IJ}$ (see \cite{mrm} for more details).

The KM algebra associated to $G_c$, denoted $\widetilde{\mathfrak{g}}(G_c)$,
is a central extension of the algebra ${\mathfrak{g}}(G_c)$. We denote $%
\mathcal{T}_{aI}$ the generators in $\widetilde{\mathfrak{g}}(G_c)$
corresponding to the generators $T_{aI}= T_a \rho_I(\varphi,\theta)$ in ${%
\mathfrak{g}}(G_c)$. Thus $\widetilde{\mathfrak{g}}(G_c) = \big\{\mathcal{T}%
_{aI},a=1\cdots, \dim \mathfrak{g}, I\in \mathcal{I},D_i,k_i, i=1,\cdots r\}
$. From \eqref{eq:hatg}, \eqref{eq:DT} and \eqref{eq:k} the Lie brackets are
\begin{eqnarray}  \label{eq:KMG}
\big[\mathcal{T}_{aI} , \mathcal{T}_{bJ}\big] &=& i f_{ab}{}^c c_{IJ}{}^K
\mathcal{T}_{cK} + \eta_{ab} \eta_{IJ} \sum \limits_{i=1}^r k_i J(i)  \notag
\\
\big[D_i,\mathcal{T}_{aI}\big] &=&I(i)\mathcal{T}_{aI} \ .
\end{eqnarray}

All the results of this section extend naturally to the coset space $G/H$,
where $H$ is a subgroup of $G$ \cite{mrm}.

\subsection{ `Softening' of (compact) Lie group manifolds}

\label{sec:def}


We consider a smooth, `softening' deformation $G_{c}^{\mu }$ of the Lie
group $G_{c}$, locally diffeomorphic to $G_{c}$ itself\footnote{%
Here, we are not going to deal with th general theory of the `softening' of
compact cosets $G_{c}/H$ (involving the so-called `horizontality condition'
for the curvatures), addressing the reader to Sec. I.3.7 of \cite%
{Castellani:1991et} for a treatment and a list of Refs..} (see e.g. \cite{cas}
for a review, and a list of Refs.). We assume that the Vielbein $\mu $ is an
intrinsic one-form (valued in the Lie algebra $\mathfrak{g}_{c}$ of $G_{c}$)
\begin{equation*}
\mu ^{A}(m)=\mu _{A}{}^{M}(m)\;\text{d}m^{M}
\end{equation*}%
\textit{i.e.}, it is not a Maurer-Cartan one-form (namely, it does not
satisfy \eqref{eq:e}). In other words, $\mu $ is \textit{not} left-invariant
(i.e., it is a `soft', intrinsic one-form), and it does \textit{not} satisfy
the Maurer-Cartan equation, but rather it holds that
\begin{equation}
\text{d}\mu +\mu \wedge \mu =R,  \label{this}
\end{equation}%
where $R$ is the curvature two-form of $\mu $. It is in this sense that we
consider that the `soft' group manifold $G_{c}^{\mu }$ is a \textit{%
deformation}\footnote{%
In the case $ISO(1,3)/SO(1,3)$, namely the coset of the Poincar\'{e} group
by the Lorentz group, the analogue of \eqref{eq:gG} in this case leads to
the flat Minkowski space-time, whilst the intrinsic definition of the
Vielbein above leads to a Riemannian space with curvature (see e.g. Sec. I.3
of \cite{Castellani:1991et}).} of $G_{c}$.

Then, $\mu $ span a basis of the cotangent plane of $G_{c}^{\mu }$, and one
can define the metric tensor
\begin{equation}
g_{MN}^{\mu }(m)=\mu _{M}{}^{A}(m)\mu _{N}{}^{B}(m)\delta _{AB}.
\label{eq:gGmu}
\end{equation}%
Taking the exterior derivative of both sides of Eq. (\ref{this}), one
obtains the Bianchi identity for the curvature of $\mu $,%
\begin{equation*}
dR+2R\wedge \mu =0\Leftrightarrow \nabla R=0,
\end{equation*}%
where the covariant derivative operator $\nabla $ on $G_{c}^{\mu }$ has been
introduced.

We will further assume that the manifold $G_{c}^{\mu }$ has the same
parameterisation $m^{M}=(\varphi ^{i},\theta ^{r})$, the only difference
between $G_{c}$ and $G_{c}^{\mu }$ being at the level of the metric tensor (%
\eqref{eq:gG} or \eqref{eq:gGmu} respectively). The corresponding scalar
product on $G_{c}^{\mu }$ is then given by
\begin{equation}
(f,g)_{\mu }=\frac{1}{V}\int_{G_{c}}\sqrt{g^{\mu }}\;\text{d}\varphi ^{p}%
\text{d}\theta ^{q}\overline{f(\varphi ,\theta )}g(\varphi ,\theta ),
\label{eq:PSGmu}
\end{equation}%
where $g^{\mu }=\det (g_{MN}^{\mu })$. Notice that, since the
parameterisation of $G_{c}^{\mu }$ and $G_{c}$ is the same, the limits of
integration are again $G_{c}$ in this case.\smallskip

Two brief remarks are in order (for further elucidation, see e.g. \cite{cas}%
).

\begin{enumerate}
\item The definition of `soft' one-forms $\mu $ and of the associated
curvature two-form $R$ is the same as in Yang-Mills theory, with the crucial
difference that in the present case, the Vielbein one-form $\mu $ is defined
on $G_{c}^{\mu }$, which does not have an \textit{apriori} fiber bundle
structure.

\item We have introduced the `soft' forms starting from the dual covariant
formulation of the Lie algebra $\mathfrak{g}_{c}$ of $G_{c}$, namely from
the Maurer-Cartan equation \eqref{eq:e}. Of course, the same can be done in
the contravariant language of vector fields.
\end{enumerate}

\subsection{A KM algebra on $G_c^ \protect\mu$}

\label{sec:KKMM}

Let $L^2(G_c^ \mu)$ be the set of square integrable functions on $G^ \mu_c$
endowed with the scalar product \eqref{eq:PSGmu}. Following Mackey \cite{Mc}%
, we can easily construct a complete set of orthonormal functions on $%
L^2(G_c^ \mu)$ (see also \cite{ram}). Recall the main points which enable to
associate a Hilbert basis on a manifold $\mathcal{M}$ with integration
measure $\text{d} \beta$ starting from a Hilbert basis on the same manifold $%
\mathcal{M}$ with integration measure $\text{d} \alpha$ (\cite{Mc}, p. 100).
These two integration measures endow $\mathcal{M}$ with two different scalar
products
\begin{eqnarray}
\begin{array}{ll}
(\mathcal{M}, \text{d} \alpha): & (f,g)_\alpha= \int_{\mathcal{M}} \text{d}%
\alpha\; \overline{f(m)} g(m) \\[.2cm]
(\mathcal{M}, \text{d} \beta): & (f,g)_\beta= \int_{\mathcal{M}} \text{d}%
\beta\; \overline{f(m)} g(m)%
\end{array}
\notag
\end{eqnarray}
with $m \in \mathcal{M}$. We assume further that there exists a mapping $%
T_{\beta \alpha}$:
\begin{eqnarray}
T_{\beta \alpha}: L^2(\mathcal{M},\text{d} \beta) \to L^2(\mathcal{M}, \text{%
d} \alpha) \ ,  \notag
\end{eqnarray}
such that
\begin{eqnarray}
\int_{\mathcal{M}} \text{d} \alpha = \int_{\mathcal{M}} \text{d} \beta
T_{\beta \alpha} . \   \notag
\end{eqnarray}
For instance, for an $n$-dimensional Riemannian manifold $\mathcal{M}$
parameterized by $m_1,\cdots,m_n$ with metric $g_\alpha$ (resp. $g_\beta$)
we have $\text{d}\alpha = \sqrt{\big|\det g_\alpha\big|} \text{d}^n m$
(resp. $\text{d}\beta = \sqrt{\big| \det g_\beta\big|} \text{d}^n m$) and
thus $T_{\beta \alpha} = \sqrt{\big|\det g_\alpha\big|/\big|\det g_\beta\big|%
}$. This means that if $\{f^\beta_i, i\in \mathbb{N}\}$ is a Hilbert basis
of $L^2(\mathcal{M},\text{d} \beta)$, then $\{f^\alpha_i = \frac{f^\beta_i }{%
\sqrt{T_{\beta \alpha}}}, i\in \mathbb{N}\}$ is a Hilbert basis for $L^2(%
\mathcal{M},\text{d} \alpha)$, and we obviously have
\begin{eqnarray}
(f^\beta_i,f^\beta_j)_\beta = \delta_{ij}\ \ \Longleftrightarrow \ \
(f^\alpha_i,f^\alpha_j)_\alpha = \delta_{ij} \   \notag
\end{eqnarray}
and the map $T_{\beta\alpha}$ is unitary.

Applied to our case, the transition functions read $T^\mu(m) = \sqrt{\frac g
{g^\mu}}$ (see \eqref{eq:PSGmu}) and the orthonormal Hilbert basis of $%
G_c^\mu$ is
\begin{eqnarray}
\mathcal{B}_\mu = \Big\{\rho_I^ \mu(\varphi,\theta) = \sqrt{T^\mu}
\rho_I(\varphi,\theta) \ , \ \ I \in \mathcal{I} \Big\},
\end{eqnarray}
trivially satisfying the relation
\begin{eqnarray}
(\rho_I^\mu,\rho^\mu_J)_\mu= \delta_{IJ} \ .  \notag
\end{eqnarray}
If we now introduce
\begin{eqnarray}
L_A^\mu=\sqrt{T^\mu} L_A \frac 1{\sqrt{T^\mu}} \ , \ \ R_A^\mu=\sqrt{T^\mu}
R_A \frac 1{\sqrt{T^\mu}} \ ,  \notag
\end{eqnarray}
it follows at once that
\begin{eqnarray}
\big[L^\mu_A,L^\mu_B\big]=i C_{AB}{}^C L^\mu_C \ , \ \ \big[R^\mu_A,R^\mu_B%
\big]=i C_{AB}{}^C R^\mu_C \ , \ \ \big[L^\mu_A,R^\mu_B\big]=0 \ .  \notag
\end{eqnarray}
and
\begin{eqnarray}  \label{eq:muRep}
L^\mu_A \rho^\mu_{LQR}(\varphi,\theta)&=& C^Q_{AL}{}^{L^{\prime
}}\rho^\mu_{L^{\prime }QR}(\varphi,\theta)  \notag \\
R^\mu_A \rho_{LQR}(\varphi,\theta)&=& C^Q_{AR}{}^{R^{\prime
}}\rho^\mu_{LQR^{\prime }}(\varphi,\theta)\
\end{eqnarray}
Thus $\{L_A^\mu, A=1,\cdots,n\}$ and $\{R_A^\mu,A=1,\cdots,n\}$ generate the
Lie algebra $\mathfrak{g}_c$ and $\rho^\mu_{LQR}$ are the corresponding
matrix elements of $G_c$ (but not of $G_c^\mu$, which is not a group).%
\newline

To define the analogue of \eqref{eq:loop} for the deformed, `soft' group
manifold $G_{c}^{\mu }$, we need more. Indeed, the products $\rho _{I}^{\mu
}\rho _{J}^{\mu }$ must be square integrable for any $I,J.$ We now show that
if for any $I,J$ the function $\rho _{I}^{\mu }\rho _{I}^{\mu }\in
L^{2}(G_{c}^{\mu })$, then $\sqrt{T^{\mu }}\in L^{2}(G_{c})$. Indeed, for
the trivial representation of $G_{c}$ it holds that $\rho _{0}(\varphi
,\theta )=1$, thus by hypothesis $(\rho _{0}^{I})^{2}=T^{\mu }\in
L^{2}(G_{c}^{\mu })$. Now $(\rho _{0}^{I})^{2}\in L^{2}(G_{c}^{\mu })$ leads
to $\sqrt{T^{\mu }}\in L^{2}(G_{c})$ because

\begin{eqnarray}
\frac 1 V \int_{G_c}\sqrt{g^ \mu}\; \text{d}\varphi^p \text{d}\theta^q
(T^\mu(\varphi,\theta))^2= \frac 1 V \int_{G_c}\sqrt{g}\; \text{d}\varphi^p
\text{d}\theta^q {T^\mu(\varphi,\theta)}
\end{eqnarray}
Consequently, $\sqrt{T^ \mu}$ is a square integrable function of $L^2(G_c)$
(we already know that indeed $\sqrt{T^ \mu}$ is a square integrable function
of $L^2(G_c^\mu)$) and we have
\begin{eqnarray}  \label{eq:T}
\sqrt{T^ \mu}= C^I \rho_I(\varphi,\theta)
\end{eqnarray}
Since now by hypothesis the product $\rho^ \mu_I\rho^ \mu_J $ belongs to $L^2(G_c^\mu)$:
\begin{eqnarray}  \label{eq:rhorhomu}
\rho^ \mu_I(\varphi, \theta)\rho^ \mu_J(\varphi, \theta)=
c^{\mu}_{IJ}{}^K \rho^ \mu_K(\varphi, \theta)
\end{eqnarray}
where, using \eqref{eq:T},
\begin{equation}
c^{\mu}_{IJ}{}^K \equiv C^L c_{IJ}^{~~M} c_{LM}^{~~K} \  \label{eq:Cmu}
\end{equation}
can be regarded as the\footnote{%
Here and below, these quotes are used to stress the slight abuse of
language, due to the fact that $G_{c}^{\mu }$ is not a group.}
`Clebsch-Gordan coefficients' of $G_{c}^{\mu }$.

Conversely, if the function $\sqrt{T^\mu}\in L^2(G_c)$ is sufficiently
smooth, then $\sqrt{T^\mu}\in L^2(G_c)$ implies that for any $I,J$, the
product $\rho^\mu_I\rho^\mu_J \in L^2(G_c^\mu)$. Indeed, (because of %
\eqref{eq:T} and \eqref{eq:rhorho}) we have
\begin{eqnarray}
\rho^\mu_I(\varphi, \theta) \rho^\mu_J(\varphi, \theta) &=& \sqrt{%
T^\mu(\varphi, \theta)} \sqrt{T^\mu(\varphi, \theta)}\rho_I(\varphi,
\theta)\rho_J(\varphi, \theta)  \notag \\
&=&\sqrt{T^\mu(\varphi, \theta)} C^K c_{IJ}{}^M \rho_K(\varphi,
\theta)\rho_M(\varphi, \theta)  \notag \\
&=& C^K c_{IJ}{}^M c_{KM}{}^N \rho^\mu_N(\varphi, \theta)  \notag \\
&=&c^\mu_{IJ}{}^N \rho_N^\mu(\varphi, \theta) \ .  \notag
\end{eqnarray}

\noindent As seen previously the $\rho _{I}^{\mu }$ are in the left and
right representation of $G_{c}$ (see \eqref{eq:muRep}). However, since the
metric tensor is deformed by the parameter $T^{\mu }$, we have to take into
account this deformation parameter when considering tensor product of
representations. In particular, we define
\begin{equation*}
\rho _{I}^{\mu }(\varphi ,\theta )\otimes _{\mu }\rho _{J}^{\mu }(\varphi
,\theta )\equiv \frac{1}{\sqrt{T^{\mu }}}\rho _{I}^{\mu }(\varphi ,\theta
)\rho _{J}^{\mu }(\varphi ,\theta )=\sqrt{T^{\mu }}\rho _{I}(\varphi ,\theta
)\rho _{J}(\varphi ,\theta )=c_{IJ}{}^{K}\rho _{K}^{\mu }(\varphi ,\theta )\
,
\end{equation*}%
with $c_{IJ}{}^{K}$ the Clebsch-Gordan coefficients of $G_{c}$, and thus
recovering the usual results (see \eqref{eq:rhorho}). Note that this
equation is very different from \eqref{eq:rhorhomu}. Indeed, in %
\eqref{eq:rhorhomu} we have considered the usual tensor product inherited
from the group $G_{c}$, whilst in the definition above we have used the
`deformed' tensor product associated to $G_{c}^{\mu }$.

Thus, under these conditions,
\begin{equation*}
\mathfrak{g}(G_{c}^{\mu })=\Big\{T_{aI}^{\mu }=T_{a}\rho _{I}^{\mu }(\varphi
,\theta )\ ,\ \ a=1,\cdots ,d\ ,\ \ I\in \mathcal{I}\Big\}\ ,
\end{equation*}%
is a Lie algebra with Lie brackets
\begin{equation*}
\big[T_{aI}^{\mu },T_{bJ}^{\mu }\big]=if_{ab}{}^{c}c_{IJ}^{\mu
}{}^{K}T_{cK}^{\mu }\ .
\end{equation*}%
The definition of Hermitian and central charges follows easily. For the
Hermitean commuting operators we introduce
\begin{equation*}
D_{i}^{\mu }=\sqrt{T^{\mu }}D_{i}\frac{1}{\sqrt{T_{\mu }}}\ ,
\end{equation*}%
which obviously leads to
\begin{equation*}
\big[D_{i}^{\mu },T_{aI}^{\mu }\big]=I(i)T_{aI}^{\mu }\ .
\end{equation*}%
Similarly, since%
\begin{equation*}
T_{aI}^{\mu }=T_{a}\rho _{I}^{\mu }=\sqrt{T^{\mu }}T_{a}\rho _{I}=\sqrt{%
T^{\mu }}T_{aI}=T_{a}C^{J}\rho _{I}\rho _{J}=T_{a}C^{J}c_{IJ}^{~~K}\rho
_{K}=C^{J}c_{IJ}^{~~K}T_{aK}=P_{I}^{~K}T_{aK},
\end{equation*}%
where\footnote{%
Such that the `Clebsch-Gordan coefficients' of $G_{c}^{\mu }$, defined by (\ref%
{eq:Cmu}), are given by $c^{\mu}_{IJ}{}^K=c_{IJ}^{~~M}P_{M}^{~K}$.}%
\begin{equation}
P_{I}^{~K} \equiv C^{J}c_{IJ}^{~~K}, \label{eq:P}
\end{equation}%
the two-cocycles are given by
\begin{equation}
\omega _{i}^{\mu }\left( T_{aI}^{\mu },T_{bJ}^{\mu }\right) =\omega
_{i}\left( P_{I}^{~I^{\prime }}T_{aI^{\prime }},P_{J}^{~J^{\prime
}}T_{bJ^{\prime }}\right) =P_{I}^{~I^{\prime }}P_{J}^{~J^{\prime }}\omega
_{i}\left( T_{aI^{\prime }},T_{bJ^{\prime }}\right) =k_{i}\eta
_{ab}P_{I}^{~I^{\prime }}P_{J}^{~J^{\prime }}J^{\prime }(i)\eta _{I^{\prime
}J^{\prime }}. \label{eq:omega}
\end{equation}%
Note that $\omega _{i}^{\mu }$
is obtained by an integration over the original manifold $G_{c}$, and not
over $G_{c}^{\mu }$. This in particular means that the differential
operators that we have to take in duality with the two-cocycles are the
original Hermitean operators $D_{i}$, and \textit{not} the operators $%
D_{i}^{\mu }$. The fact that $\omega _{i}$ are two-cocycles ensure
naturally that $\omega _{i}^{\mu }$ are also two-cocycles. Indeed, both
cocycles are defined by the same integration upon $G_{c}$.

The KM algebra associated to $G^\mu_c$ is given by $\widetilde{\mathfrak{g}}%
(G^ \mu_c) = \big\{\mathcal{T}^ \mu_{aI} \ ,\ \ a = 1,\cdots, d \ , I
\in \mathcal{I}, D_i,k_i, i=1,\cdots r\}$, where $\mathcal{T}_{aI}^\mu$
are the generators in $\widetilde{\mathfrak{g}}(G^ \mu_c)$ corresponding to
the generators $T_{aI}^\mu=T_a \rho_I^\mu(\varphi,\theta)$ in $\mathfrak{g}%
(G_c^\mu)$. The Lie brackets read\footnote{%
Despite having adopted the Einstein summation convention on repeated
indices, here we keep the $\sum_{i=1}^{r}$ symbol to stress the sum over all
$r$ central extensions.}
\begin{eqnarray}
\left[ \mathcal{T}_{aI}^{\mu },\mathcal{T}_{bJ}^{\mu }\right]
&=&if_{ab}^{~~c}c^{\mu}_{IJ}{}^K\mathcal{T}_{cK}^{\mu }+\sum_{i=1}^{r}\omega
_{i}^{\mu }\left( \mathcal{T}_{aI}^{\mu },\mathcal{T}_{bJ}^{\mu }\right) =
\nonumber \\
&=&if_{ab}^{~~c}c^{\mu}_{IJ}{}^K\mathcal{T}_{cK}^{\mu }+P_{I}^{~I^{\prime
}}P_{J}^{~J^{\prime }}\eta _{ab}\eta _{I^{\prime }J^{\prime
}}\sum_{i=1}^{r}k_{i}J^{\prime }(i);  \nonumber \\
\left[ D_{i},\mathcal{T}_{aI}^{\mu }\right]  &=&P_{I}^{~I^{\prime
}}I^{\prime }(i)\mathcal{T}_{aI^{\prime }}^{\mu }.\label{eq:KMGmu}
\end{eqnarray}

Even if the KM algebra associated to $G_{c}$ and the KM associated to $%
G_{c}^{\mu }$ seem to be very similar (see \eqref{eq:KMG} and %
\eqref{eq:KMGmu}), they have important structural differences. In the first
case, the Hilbert basis of $L^{2}(G_{c})$ is obtained by the matrix elements
of (finite-dimensional) unitary representations of $G_{c}$ (Peter--Weyl
theorem), while in the second case, the Hilbert basis of $L^{2}(G_{c}^{\mu
}) $ is also related to representation theory of $G_{c}$, but not of $%
G_{c}^{\mu }$, which generally does not have a group structure. This can be
regarded as a consequence of the \textit{principle of equivalence} of general
relativity, holding within the manifold $G_{c}^{\mu }$, whose systems of
flat coordinates pertain to $G_{c}$. Furthermore, the manifold $G_{c}$
possesses an obvious isometry, corresponding to the action of $G_{c}$ onto
itself, whereas the manifold $G_{c}^{\mu }$ does not admit isometries in
general. For the same reasons, it is irrelevant whether the
`softening' procedure is applied to the Lie algebra $\mathfrak{g}$ (namely, to $%
\mathfrak{g}(G_{c})$, $\mathfrak{g}(G_{c}^{\mu })$, or to central extension
thereof, $\widetilde{\mathfrak{g}}(G_{c}^{\mu })$) : at the Lie algebra
level, i.e. locally on the corresponding group manifold $\mathcal{G}$ (such
that Lie$(\mathcal{G})=\mathfrak{g})$, the `softening' has no non-trivial
action, and thus, trivially\footnote{%
For simplicity's sake, here we assume $\mathfrak{g}$ to be compact; see \eqref{eq:g}.} : $\mathfrak{%
g}^{\mu }(G_{c})\equiv \mathfrak{g}(G_{c})$, and\footnote{%
The priming of the lowercase Greek indices denotes two (\textit{a priori}
different and independent) `softening' procedures.} $\mathfrak{g}^{\mu
}\left( G_{c}^{\mu ^{\prime }}\right) \equiv \mathfrak{g}(G_{c}^{\mu
^{\prime }})$.

\section{`Softening' of $\mathbb{S}^{1}$ and the associated KM algebra}

\label{sec:defS1} As a first example, let us study a KM algebra associated
to the `softened' deformation of the one-sphere (circle) $\mathbb{S}^{1}$.
The KM algebra associated to the undeformed circle $\mathbb{S}^{1}$ is
nothing but the affine Lie algebra $\widetilde{\mathfrak{g}}$.

The Hilbert space $L^2(\mathbb{S}^1)$ is endowed with the natural scalar
product:
\begin{eqnarray}  \label{eq:S1}
(f,g)= \frac 1{2\pi} \int \limits_{0}^{2\pi} \text{d}\theta\;\overline{%
f(\theta)} g(\theta) \ ,
\end{eqnarray}
and the Hilbert basis is
\begin{eqnarray}
\mathcal{B}=\big\{ e_n(\theta) = e^{in\theta}\ , \ \ n\in \mathbb{Z}\big\} \
.  \notag
\end{eqnarray}
Since%
\begin{equation*}
e_{n}\left( \theta \right) e_{m}\left( \theta \right) =e_{n+m}\left( \theta
\right) ,
\end{equation*}
the Clebsch-Gordan coefficients of $\mathbb{S}^{1}$ are simply given by%
\begin{equation}
c_{mn}^{~~p}=\delta _{p,m+n}.\label{CG-S1}
\end{equation}

Let $d=-i\partial_\theta$, and $\mathfrak{g}({\mathbb{S}^1%
})=\{T_{am}=T_a e_m(\theta)\}$ be the loop algebra associated to $\mathfrak{g%
}$. The central extension is associated to the two-cocycle:
\begin{eqnarray}
\omega(T_{am},T_{bn}) = \frac k{2\pi} \int \limits_0^{2\pi} \text{d}\theta\;%
\Big<T_{am}, d T_{bn}\Big>_0= kn \;\eta_{ab}\delta_{n,-m} \ .  \notag
\end{eqnarray}
Thus, the affine Lie algebra (which centrally extends the loop algebra) is
\begin{eqnarray}
\widetilde{\mathfrak{g}}=\Big\{\mathcal{T}_{an}, k, d, a=1,\cdots,d, m \in
\mathbb{Z}\Big\}  \notag
\end{eqnarray}
and the Lie brackets are given by
\begin{eqnarray}  \label{eq:aff}
\big[\mathcal{T}_{am},\mathcal{T}_{bn}\big] &=& if_{ab}{}^c \mathcal{T}_{c|
m+n} + kn\eta_{ab}\delta_{m,-n}  \notag \\
\big[d,\mathcal{T}_{am}\big]&=& m \mathcal{T}_{am} \ .
\end{eqnarray}

Let $f$ and $g$ be two functions on the circle. The algebra of vector fields
(i.e., the de Witt algebra) is given by the Lie brackets
\begin{eqnarray}
\big[f(\theta)\partial_\theta, g(\theta)\partial_\theta\big]= \Big(f(\theta)
g^{\prime }(\theta) - g(\theta)f^{\prime }(\theta)\Big) \partial_\theta \ .
\notag
\end{eqnarray}
A natural basis of the de Witt algebra is given by $\{\ell_n, n\in \mathbb{Z}%
\}$ with $\ell_n(\theta) = i e^{in\theta}\partial_\theta$. Over this basis,
the brackets acquire the form:
\begin{eqnarray}  \label{eq:dW}
\big[\ell_m,\ell_n\big]=(m-n)\ell_{m+n} \ .
\end{eqnarray}
The de Witt algebra admits a central charge given by the Gel'fand--Fuks
cocycle \cite{Fuks}
\begin{eqnarray}
\omega(f,g)=-i\frac c{12}\frac 1{2\pi} \int \limits_0^{2\pi} \text{d}%
\theta\; f(\theta)g^{\prime \prime \prime }(\theta)
\end{eqnarray}
In particular, it follows that
\begin{eqnarray}
\omega(\ell_m,\ell_n)= \frac{c}{12}m^3 \delta_{m+n,0} \ .
\end{eqnarray}
Let $\mathcal{L}_m,m\in \mathbb{Z}$ be the generators of the Virasoro
algebra, \textit{i.e.}, the centrally extended de Witt algebra. The Lie
brackets reduce to
\begin{eqnarray}
\big[\mathcal{L}_m,\mathcal{L}_n\big]=(m-n)\mathcal{L}_{m+n} + \frac c{12}
m^3 \delta_{m+n,0} \ .  \notag
\end{eqnarray}
A more convenient basis is given by $L_m= \mathcal{L}_m+\frac
c{24}\delta_{n,0}$. Over this basis, we obtain the standard Lie brackets of
the Virasoro algebra:
\begin{eqnarray}  \label{eq:Vir}
\big[L_m,L_n\big]=(m-n) L_{m+n} + \frac c{12}(m^3-m)\delta_{m+n,0} \ .
\end{eqnarray}
The Virasoro algebra \eqref{eq:Vir} and the affine Lie algebra \eqref{eq:aff}
admit a semi-direct structure:
\begin{eqnarray}  \label{eq:VirAff}
\big[L_m, T_{an}\big]= -n T_{a n+m} \ ,
\end{eqnarray}
thus the Hermitean operator $d$ of the affine Lie algebra can be identified
with $-L_0$.\newline

To construct a deformation of \eqref{eq:aff}, following Sections \ref{sec:def} and \ref{sec:KKMM}
, we introduce the deformed scalar product on the `soft' circle $\mathbb{S}%
_{F}^{1}$:
\begin{equation}
(f,g)_{F}=\frac{1}{2\pi }\int\limits_{0}^{2\pi }F(\theta )\text{d}\theta \;%
\overline{f(\theta )}g(\theta )\ , \label{eq:spF}
\end{equation}%
where $F$ is a positive function such that $1/\sqrt{F}$ belongs to $L^{2}(%
\mathbb{S}^{1})$. The Hilbert basis of $L^{2}(\mathbb{S}_{F}^{1})$ is
\begin{equation}
\mathcal{B}_{F}=\Big\{e_{n}^{F}(\theta )=\frac{1}{\sqrt{F(\theta )}}%
e^{in\theta }\ ,\ \ n\in \mathbb{Z}\Big\} \label{eq:BF}
\end{equation}%
(the set of functions $\{e_{n}^{F}\}$ constitutes an orthonormal set with respect to the scalar product \eqref{eq:spF}). Since\footnote{%
To be very explicit, in this Section, we do not use Einstein's summation
convention on repeated indices.}%
\begin{equation*}
\frac{1}{\sqrt{F\left( \theta \right) }}=\sum_{n\in \mathbb{Z}%
}F^{n}e_{n}\left( \theta \right) =\sum_{n\in \mathbb{Z}}F^{n}e^{in\theta },
\end{equation*}
the `Clebsch-Gordan coefficients' of $\mathbb{S}_{F}^{1}$ are given by%
\begin{equation*}
c_{mn}^{F~~p}=\sum_{k,q\in \mathbb{Z}}F^{k}c_{mn}^{~~q}c_{kq}^{~~p}=%
\sum_{k,q\in \mathbb{Z}}F^{k}\delta _{q,m+n}\delta _{p,k+q}=\sum_{k\in
\mathbb{Z}}F^{k}\delta _{p,m+n+k}=F^{p-m-n},
\end{equation*}
whereas the coefficients $P_{n}^{~n^{\prime }}$, generally defined by (\ref%
{eq:P}), are in this case given by%
\begin{equation*}
P_{n}^{~n^{\prime }}=\sum_{m\in \mathbb{Z}}F^{m}c_{mn}^{~~n^{\prime
}}=\sum_{m\in \mathbb{Z}}F^{m}\delta _{n^{\prime },m+n}=F^{n^{\prime }-n},
\end{equation*}
such that%
\begin{equation*}
c_{mn}^{F~~p}=P_{m+n}^{~~~p}.
\end{equation*}.
Moreover, the Hilbert basis of $L^{2}\left( \mathbb{S}_{F}^{1}\right) $ \eqref{eq:BF} can be
rewritten as%
\begin{equation*}
\mathcal{B}_{F}=\left\{ e_{n}^{F}\left( \theta \right) =\sum_{n^{\prime }\in
\mathbb{Z}}P_{n}^{~n^{\prime }}e_{n^{\prime }}\left( \theta \right) ,~~n\in
\mathbb{Z}\right\} .
\end{equation*}
As in the general treatment of Section \ref{sec:KKMM}, the natural choice
for the Hermitean operator associated to $\mathbb{S}_{F}^{1}$ would be
\begin{equation}  \label{eq:dF}
d_{F}=-i\frac{1}{\sqrt{F(\theta )}}\partial _{\theta }\sqrt{F(\theta )}\ ,
\end{equation}%
but, as treated above, the differential operator that we have to take in
duality with the two-cocycle $\omega ^{F}$ is the original Hermitean
operator $d$, and not the operator $d_{F}$.

By observing that the generators of $\mathfrak{g}\left( \mathbb{S}%
_{F}^{1}\right) $ are $T_{am}^{F}\equiv \frac{1}{\sqrt{F\left( \theta
\right) }}T_{am}$, the two-cocycle of $\mathbb{S}_{F}^{1}$ reads%
\begin{eqnarray}
\omega _{F}\left( T_{am}^{F},T_{bn}^{F}\right)  &=&\sum_{m^{\prime
},n^{\prime }\in \mathbb{Z}}P_{m}^{~m^{\prime }}P_{n}^{~n^{\prime }}\omega
\left( T_{am^{\prime }},T_{bn^{\prime }}\right) =k\eta _{ab}\sum_{m^{\prime
},n^{\prime }\in \mathbb{Z}}n^{\prime }\delta _{n^{\prime },-m^{\prime
}}F^{m^{\prime }-m}F^{n^{\prime }-n}  \nonumber \\
&=&k\eta _{ab}\sum_{n^{\prime }\in \mathbb{Z}}n^{\prime }F^{-n^{\prime
}-m}F^{n^{\prime }-n}.
\end{eqnarray}%

Thus, from Sections \ref{sec:def} and \ref{sec:KKMM} the corresponding deformation of the affine
Lie algebra is given by
\begin{equation*}
\widetilde{\mathfrak{g}}(\mathbb{S}_{F}^{1})=\Big\{\mathcal{T}%
_{am}^{F},\;d,\;k\Big\},
\end{equation*}
with Lie brackets reading
\begin{eqnarray}
\left[ \mathcal{T}_{am}^{F},\mathcal{T}_{bn}^{F}\right]
&=&if_{ab}^{~~c}\sum_{p\in \mathbb{Z}}c_{mn}^{F~~p}\mathcal{T}%
_{cp}^{F}+\omega _{F}\left( \mathcal{T}_{am}^{F},\mathcal{T}_{bn}^{F}\right)
=  \nonumber \\
&=&if_{ab}^{~~c}\sum_{p\in \mathbb{Z}}F^{p}\mathcal{T}_{c|m+n+p}^{F}+k\eta
_{ab}\sum_{n^{\prime }\in \mathbb{Z}}n^{\prime }F^{-n^{\prime
}-m}F^{n^{\prime }-n};  \nonumber \\
\left[ d,\mathcal{T}_{am}^{F}\right]  &=&\sum_{m^{\prime }\in \mathbb{Z}%
}P_{m}^{~m^{\prime }}m^{\prime }\mathcal{T}_{am^{\prime }}^{F}=\sum_{p\in
\mathbb{Z}}\left( m+p\right) F^{p}\mathcal{T}_{a|m+p}^{F}.\label{eq:KMS1F}
\end{eqnarray}

Next, we would like to construct a central extension of the algebra of
vector fields on $\mathbb{S}_{F}^{1}$. It is natural to consider $\ell
_{m}^{F}(\theta )=\frac{i}{\sqrt{F(\theta )}}e^{im\theta }\partial _{\theta }%
\sqrt{F(\theta )}$ (see $d_{F}$ in \eqref{eq:dF}). However, in this case the
Lie bracket of the vectors $[\ell _{m}^{F},\ell _{n}^{F}]$ on $\mathbb{S}%
_{F}^{1}$ coincide with the Lie bracket of the de Witt algebra \eqref{eq:dW}%
. This in particular means that the Virasoro algebra associated to the
`soft' circle reduces to the usual Virasoro algebra. In other words, we can
also see that the Gel'fand--Fuks cocycle $\tilde{\omega}^{F}$ associated to $\{\ell _{n}^{F},n\in
\mathbb{\}}$ is given by
\begin{equation}
\tilde{\omega}^{F}(e_{n}^{F},e_{m}^{F})=-i\frac{c}{12}\frac{1}{2\pi }\int
F(\theta )\text{d}\theta \;e_{n}^{F}(\theta )d_{F}^{3}e_{m}^{F}(\theta
)=\omega (e_{n},e_{m})\ .  \label{eq:S1F}
\end{equation}%
We conclude that on the `soft' circle we can define a semi-direct product of
the deformed affine Lie algebra \eqref{eq:KMS1F} with the Virasoro algebra
(see \eqref{eq:Vir} and \eqref{eq:VirAff}).

The curvature associated to the metric of the `soft' circle is $R=0$. Thus
we can perform a global change of variable in \eqref{eq:S1F} $\theta
\rightarrow \psi $ such that
\begin{equation*}
F(\theta )=\frac{\text{d}\psi }{\text{d}\theta }\ ,
\end{equation*}%
and the scalar product \eqref{eq:S1F} reduces within the new coordinate $%
\psi $ to the scalar product on the undeformed circle \eqref{eq:S1}.

This in particular means that the semi-direct product of the affine Lie
algebra with the Virasoro algebra associated to the `soft' circle is
trivially isomorphic to its undeformed analogue. This is not surprising,
since the topological classification of one-dimensional closed curves shows
that all such curves are topologically equivalent to the circle $\mathbb{S}%
^{1}$. This equivalence can be formalized by stating that any closed,
connected, one-dimensional manifold without boundary is homeomorphic to $%
\mathbb{S}^{1}$. Specifically, a homeomorphism is a continuous bijection
with a continuous inverse, and in the case of one-dimensional closed curves,
this means that regardless of how the curve is deformed or embedded in
space, it retains the same topological structure as $\mathbb{S}^{1}$.
Mathematically, this follows from the fact that the classification of
one-dimensional manifolds shows that $\mathbb{S}^{1}$ is the only connected,
compact, boundaryless one-dimensional manifold (cf. e.g. \cite{Munkres2000}%
). This result is critical in applications across various fields of
mathematics and physics. For instance, in knot theory, while different knots
(which are embeddings of $\mathbb{S}^{1}$ into three-dimensional space) may
not be equivalent in terms of their embeddings, topologically all knots are
still homeomorphic to $\mathbb{S}^{1}$. This is because the study of knots
focuses on the way $\mathbb{S}^{1}$ is embedded in $\mathbb{R}^{3}$, but the
fundamental topological nature of the curve remains the same. Similarly, in
string theory, the worldsheet of a closed string is topologically equivalent
to $\mathbb{S}^{1}\times \mathbb{R}$, where $\mathbb{S}^{1}$ represents the
closed loop of the string at any fixed point in time. The universal property
of $\mathbb{S}^{1}$ as the fundamental one-dimensional closed manifold
simplifies the analysis of string propagation and interactions (see e.g.
\cite{Green1987}). Thus, the circle $\mathbb{S}^{1}$ serves as the canonical
model for all one-dimensional closed curves in topology. Of course, this
means that there is no need to define analogues of the affine Lie or
Virasoro algebras, as we have just seen.

\section{`Softening' of $\mathbb{S}^{3}$ and the associated KM algebra}

As a second example, we consider now the KM algebra associated to the
`softening'of the three-sphere $\mathbb{S}^{3}$. We first reproduce the
construction of the KM algebra associated to $SU(2)=\mathbb{S}^{3}$ proposed
in \cite{mrm}. As we want to present these results in an explicit way, we
reconsider the construction of $\widetilde{\mathfrak{g}}($SU$(2))$ in some
detail, using a different approach to that adopted in \cite{mrm}, but along
the lines of \cite{ram}.

\subsection{The three-sphere $\mathbb{S}^3$}

The group SU$(2)$ is defined by the set of $2 \times 2$ complex matrices
\begin{eqnarray}  \label{eq:SU2}
SU(2)=\Bigg\{U =
\begin{pmatrix}
z_1 & -\bar z_2 \\
z_2 & \phantom{-}\bar z_1%
\end{pmatrix}
: \ \ z_1, z_2 \in \mathbb{C}\ , \ \ |z_1|^2 + |z_2|^2 =1 \Bigg\} \cong
\mathbb{S}^3
\end{eqnarray}
The Lie algebra $\mathfrak{su}(2)$ of $SU(2)$ is generated by $J_0,J_\pm$
with Lie brackets:
\begin{eqnarray}
\big[J_0,J_\pm\big]= \pm J_\pm \ , \ \ \big[J_+,J_-\big] = 2 J_0 \ .  \notag
\end{eqnarray}

A parameterisation of the three-sphere $\mathbb{S}^{3}$ is given by
\begin{equation*}
\begin{split}
z_{1}=& \cos \frac{\theta }{2}e^{i\frac{\varphi +\psi }{2}} \\
z_{2}=& \sin \frac{\theta }{2}e^{i\frac{\varphi -\psi }{2}}
\end{split}%
\end{equation*}%
with\footnote{%
The asymmetry between the $\varphi $ and $\psi $ is resolved by Bargmann
\cite{bar} (p. 596 eq(4.15)) since he considers $-2\pi \leq \varphi ,\psi
<2\pi $, but this parameterisation covers twice the three-sphere.}
\begin{equation*}
0\leq \theta \leq \pi \ ,\ \ 0\leq \varphi <2\pi \ ,\ \ 0\leq \psi <4\pi \ .
\end{equation*}%
The left/right invariant vectors fields (obtained by solving %
\eqref{eq:Killing}) are given by
\begin{equation*}
\begin{array}{llllll}
L_{\pm } & = & e^{\pm i\psi }\Big(-\frac{i}{\sin \theta }\partial _{\varphi
}+i\cot \theta \partial _{\psi }\pm \partial _{\theta }\Big)\ , & L_{0} & =
& -i\partial _{\psi }\nonumber \\
R_{\pm } & = & e^{\pm i\varphi }\Big(\phantom{-}\;\frac{i}{\sin \theta }%
\partial _{\psi }-i\cot \theta \partial _{\varphi }\mp \partial _{\theta }%
\Big)\ , & R_{0} & = & -i\partial _{\varphi }\nonumber%
\end{array}%
\end{equation*}%
and satisfy the commutation relations
\begin{equation*}
\begin{array}{llllll}
\big[L_{0},L_{\pm }\big] & = & \pm L_{\pm }\ , & \big[L_{+},L_{-}\big] & = &
2L_{0} \\[2pt]
\big[R_{0},R_{\pm }\big] & = & \pm R_{\pm }\ , & \big[R_{+},R_{-}\big] & = &
2R_{0}%
\end{array}%
\end{equation*}%
as well as
\begin{equation*}
\big[L_{a},R_{b}\big]=0\ .
\end{equation*}%
Observe that $R_{\pm }$ and $L_{\pm }$ are related by means of
\begin{equation*}
e^{\mp i\psi }L_{\pm }+e^{\mp i\varphi }R_{\pm }=0,
\end{equation*}%
implying that the corresponding left- and right-invariant one-forms will
have the same structure, by replacing $\varphi $ by $\psi $ (see Eqs.[\ref%
{eq:RI1F}] and [\ref{eq:LI1F}]). The Casimir operator takes the form
\begin{equation*}
Q=-\partial _{\theta }^{2}-\cot \theta \partial _{\theta }-\frac{1}{\sin
^{2}\theta }\Big(\partial _{\varphi }^{2}+\partial _{\psi }^{2}\Big)+2\frac{%
\cos \theta }{\sin ^{2}\theta }\partial _{\varphi }\partial _{\psi }\ .
\end{equation*}

The $SU(2)$ right-invariant one-forms (see Eq.[\ref{eq:inv1form}]) read
\begin{eqnarray}  \label{eq:RI1F}
\omega_1&=& \sin \psi \text{d} \theta -\cos\psi \sin\theta \text{d} \varphi
\notag \\
\omega_2&=& \cos \psi \text{d} \theta +\sin\psi \sin\theta \text{d} \varphi
\\
\omega_3&=&\text{d} \psi + \cos \theta \text{d} \varphi,  \notag
\end{eqnarray}
and satisfy $d\omega _{i}=\mathbf{-} \epsilon^{jk}{}_i\omega _{j}\wedge
\omega _{k}$ ($i,j,k=1,2,3$ with summation over repeated indices
understood), which are the Maurer-Cartan equations (their right-invariance
with respect to $SU(2)$ is discussed in App. \ref{sec:App}).

The $SU(2)$ right-invariant one-forms \eqref{eq:RI1F} enable to define the
metric on the round three-sphere:
\begin{eqnarray}
\text{d}s^{2} &=&\omega _{1}^{2}+\omega _{2}^{2}+\omega _{3}^{2}  \notag \\
&=&\text{d}\theta ^{2}+\text{d}\psi ^{2}+\text{d}\varphi ^{2}+2\cos \theta
\text{d}\psi \text{d}\varphi \ .  \notag
\end{eqnarray}%
Note also that the one-form $\omega _{3}$ is invariant under $L_{0}$ and
that
\begin{eqnarray}
\lambda _{1} &=&\sin \varphi \text{d}\theta -\cos \varphi \sin \theta \text{d%
}\psi  \notag  \label{eq:LI1F} \\
\lambda _{2} &=&\cos \varphi \text{d}\theta +\sin \varphi \sin \theta \text{d%
}\psi \\
\lambda _{3} &=&\cos \theta \text{d}\psi +\text{d}\varphi  \notag
\end{eqnarray}%
are left-invariant one-forms, and $\lambda _{3}$ is invariant under $R_{0}$.
Of course, we have
\begin{equation*}
\text{d}s^{2}=\lambda _{1}^{2}+\lambda _{2}^{2}+\lambda _{3}^{2}=\omega
_{1}^{2}+\omega _{2}^{2}+\omega _{3}^{2}\ ,
\end{equation*}%
and thus the metric is left and right invariant, as expected by compactness,
reproducing the results above.

Introducing the metric tensor
\begin{eqnarray}
g=
\begin{pmatrix}
1 & 0 & 0 \\
0 & 1 & \cos\theta \\
0 & \cos \theta & 1%
\end{pmatrix}
\notag
\end{eqnarray}
we define the (normalized) invariant measure of integration on the
three-sphere:
\begin{eqnarray}  \label{eq:int}
\int \limits_{\mathbb{S}^3} \text{d} \mu(\mathrm{SU}(2))=\frac 1 {16
\pi^2}\int_{\mathbb{S}^3} \sqrt{\det g} \;\text{d} \theta \;\text{d}
\varphi\; \text{d}\psi =\frac 1 {16 \pi^2} \int \limits_{0}^{\pi} \sin
\theta \;\text{d} \theta \int \limits_{0}^{2\pi} \text{d} \varphi \int
\limits_{0}^{4\pi} \text{d} \psi \ .
\end{eqnarray}

\subsubsection{Matrix elements}

Recall that the unitary representations of SU$(2)$ are given by $\mathcal{D}%
_s = \big\{\left|s,n\right>, -s \le n\le s \big\}, s \in \frac 12 \mathbb{N}
$ and that we have
\begin{eqnarray}  \label{eq:suAction}
J_\pm \big|s,n\big>&=& \sqrt{(s\mp n)(s\pm n+1)} \big|s,n\pm 1\big>  \notag
\\
J_0 \big|s,n\big>&=&n\big|s,n\big> \\
Q \big|s,n\big>&=&s(s+1) \big|s,n\big>  \notag
\end{eqnarray}
To compute the normalized matrix elements $\psi_{nsm}, -s\le n,m \le s$ of
the representation $\mathcal{D}_s$ we proceed as in \cite{ram}.

\begin{enumerate}
\item We solve the differential equations
\begin{eqnarray}
L_0 \psi_{nsm}(\theta,\varphi,\psi)&=&n \psi_{nsm}(\theta,\varphi,\psi)
\notag \\
R_0 \psi_{nsm}(\theta,\varphi,\psi)&=&m \psi_{nsm}(\theta,\varphi,\psi)
\notag \\
Q \psi_{nsm}(\theta,\varphi,\psi)&=&s(s+1) \psi_{nsm}(\theta,\varphi,\psi) \
.  \notag
\end{eqnarray}
The first two equations lead obviously to
\begin{eqnarray}
\psi_{nsm}(\theta,\varphi,\psi)&=& e^{i n \psi + im \varphi} F_{nsm}(\theta)
\ .  \notag
\end{eqnarray}
The last equation is solved using an appropriate ansatz which depends on the
value of $m,n$ (see below and \cite{ram}) and expresses the matrix elements
in terms of hypergeometric polynomials. At this point the matrix elements
are defined up to a constant $C_{nsm}$.

\item We impose that the matrix elements satisfy both for the left and the
right action of the Lie algebra $\mathfrak{su}(2)$ relations %
\eqref{eq:suAction}. This fixes the constant $C_{nsm}$ up to a global factor
$C_s$.

\item Using the integration \eqref{eq:int} we impose the normalization
condition
\begin{eqnarray}
\big\| \psi_{nsm}\big\|^2 =\int \limits_{\mathbb{S}^3} \text{d} \mu(\mathrm{%
SU}(2)) \;\big|\psi_{nsm}(\theta,\varphi,\psi)\big|^2 = 1 \ .  \notag
\end{eqnarray}
We recall that we have for $a,b \in \mathbb{N}$
\begin{eqnarray}
\int\limits_0^\pi \text{d} \theta \; \sin \theta \cos^{2a} \frac\theta 2
\sin^{2b} \frac \theta 2 = 2 \frac{ a! b !}{(a+b+1)!} \ .  \notag
\end{eqnarray}
\end{enumerate}

We thus obtain
\begin{eqnarray}
\psi_{nsm}(\theta,\varphi,\psi) = \left\{
\begin{array}{lc}
\frac {(-1)^{m-n} \sqrt{(2s+1)}} {(n-m)!}\sqrt{ \frac{(s+n)!}{(s-n)!} \frac{%
(s-m)!}{(s+m)!}} e^{im\varphi + i n \psi} \cos^{-n-m} \frac \theta 2
\sin^{n-m} \frac \theta 2 & n\ge m\  \\
{}_2F_1(-m-s,-m+s+1;1-m+n; \; \sin^2 \frac \theta 2) & -n-m\ge 0 \\[6pt]
\frac { \sqrt{(2s+1)}} {(m-n)!} \sqrt{ \frac{(s+m)!}{(s-m)!} \frac{(s-n)!}{%
(s+n)!}} e^{im\varphi + i n \psi} \cos^{-n-m} \frac \theta 2 \sin^{m-n}
\frac \theta 2 & m\ge n\  \\
{}_2F_1(-n-s,-n+s+1;1-n+m; \; \sin^2 \frac \theta 2) & -n-m\ge 0 \\[6pt]
\frac {(-1)^{m-n} \sqrt{2s+1}} {(n-m)!} \sqrt{ \frac{(s+n)!}{(s-n)!} \frac{%
(s-m)!}{(s+m)!}} e^{im\varphi + i n \psi} \cos^{n+m} \frac \theta 2
\sin^{n-m} \frac \theta 2 & n\ge m\  \\
{}_2F_1(n-s,n+s+1;1+n-m; \; \sin^2 \frac \theta 2) & n+m\ge 0 \\[6pt]
\frac{\sqrt{2s+1}} {(m-n)!} \sqrt{ \frac{(s+m)!}{(s-m)!} \frac{(s-n)!}{(s+n)!%
}} e^{im\varphi + i n \psi} \cos^{n+m} \frac \theta 2 \sin^{m-n} \frac
\theta 2 & m\ge n\  \\
{}_2F_1(m-s,m+s+1;1+m-n; \; \sin^2 \frac \theta 2) & n+m\ge 0%
\end{array}
\right.  \notag
\end{eqnarray}
where ${}_2F_1$ denotes the Euler hypergeometric polynomial (see e.g. \cite%
{ram} for definition). Under these conditions, the set
\begin{eqnarray}  \label{eq:HS3}
\mathcal{B} = \Big\{\psi_{nsm}, s \in \frac 12 \mathbb{N}, -s\le n,m \le s %
\Big\}
\end{eqnarray}
constitutes a Hilbert basis of $L^2(\mathrm{SU}(2))$ and we have:
\begin{eqnarray}
(\psi_{nsm},\psi_{n^{\prime },s^{\prime },m^{\prime }}) = \int \limits_{%
\mathbb{S}^3} \text{d}\mu(\mathrm{SU}(2))\; \overline{\psi_{nsm}}%
(\theta,\varphi,\psi) \psi_{n^{\prime }s^{\prime }m^{\prime
}}(\theta,\varphi,\psi)= \delta_{ss^{\prime }} \delta_{n n^{\prime }}
\delta_{mm^{\prime }} \ .  \notag
\end{eqnarray}

\subsubsection{Clebsch-Gordan coefficients}

\label{sec:CC}

Considering the tensor product
\begin{eqnarray}
\mathcal{D}_{s_1} \otimes \mathcal{D}_{s_2} = \bigoplus
\limits_{S=|s_1-s_2|}^{s_1+s_2} \mathcal{D}_{S}\ ,  \notag
\end{eqnarray}
and introducing the Clebsch-Gordan coefficients $%
\begin{pmatrix}
s_1 & s_2 & S \\
m_1 & m_2 & m_1+m_2%
\end{pmatrix}%
$ we have
\begin{eqnarray}  \label{eq:C}
\big|S,m_1+m_2\big>= \sum\limits _{S=|s_1-s_2|}^{s_1+s_2} {\scriptsize
\begin{pmatrix}
s_1 & s_2 & S \\
m_1 & m_2 & m_1+m_2%
\end{pmatrix}
\mid s_1,m_1\rangle \otimes \mid s_2, m_2\rangle \ .}  \notag
\end{eqnarray}
Since our matrix elements are normalized such that
\begin{eqnarray}
\psi_{nsm}(0,0,0)= \sqrt{2s+1}  \notag
\end{eqnarray}
we obtain
\begin{eqnarray}
\psi_{n_1s_1m_1}(\theta,\varphi,\psi) \psi_{n_2s_2m_2}(\theta,\varphi,\psi)
= \sum \limits_{S=|s_1-s_2|}^{s_1+s_2} c_{s_1 s_2}^{S}{}_{n_1n_2 m_1m_2}\;
\psi_{n_1+n_2Sm_1+m_2}(\theta,\varphi,\psi)  \notag
\end{eqnarray}
with
\begin{eqnarray}  \label{eq:CCSU2}
c_{s_1 s_2}^{S}{}_{n_1n_2 m_1m_2} \equiv \sqrt{ \frac{(2s_1+1)(2s_2+1)}{2S+1}}
\begin{pmatrix}
s_1 & s_2 & S \\
n_1 & n_2 & n_1+n_2%
\end{pmatrix}%
\; \overline{
\begin{pmatrix}
s_1 & s_2 & S \\
m_1 & m_2 & m_1+m_2%
\end{pmatrix}%
}
\end{eqnarray}

\subsection{A KM algebra associated to $\mathbb{S}^3$}

\label{sec:KMG} The KM algebra associated to SU$(2)=\mathbb{S}^{3}$ follows
directly from \eqref{eq:CCSU2} and \cite{mrm}. Indeed we have
\begin{equation*}
\widetilde{\mathfrak{g}}(\text{SU}(2))=\Big\{\mathcal{T}%
_{ansm},L_{0},R_{0},k_{L},k_{R},\ \ a=1,\cdots ,d\ ,\ell \in \frac{1}{2}%
\mathbb{N},-\ell \leq n,m\leq \ell \Big\}
\end{equation*}%
with the central charges $k_{L},k_{R}$ associated to the two-cocycles:
\begin{eqnarray}
\omega _{L}(x,y) &=&k_{L}\int\limits_{\mathbb{S}^{3}}\text{d}\mu (\mathrm{SU}%
(2))\;\big<x,L_{0}y\big>_{0}  \notag \\
\omega _{R}(x,y) &=&k_{R}\int\limits_{\mathbb{S}^{3}}\text{d}\mu (\mathrm{SU}%
(2))\;\big<x,R_{0}y\big>_{0}  \notag
\end{eqnarray}%
The Lie brackets take then the form (see \cite{mrm})
\begin{eqnarray}
\big[\mathcal{T}_{ansm},\mathcal{T}_{a^{\prime }n^{\prime }s^{\prime
}m^{\prime }}\big] &=&if_{aa^{\prime }}{}^{a^{\prime \prime }}c_{ss^{\prime
}nn^{\prime }mm^{\prime }}^{s^{\prime \prime }}\mathcal{T}_{a^{\prime \prime
}n+n^{\prime }s^{\prime \prime }m+m^{\prime }}  \notag \\
&&+\eta _{ab}(-1)^{m-n}\delta _{ss^{\prime }}\delta _{n,-n^{\prime
}}\delta _{m,-m^{\prime }}(k_{L}n^{\prime }+k_{R}m^{\prime })  \notag \\
\big[L_{0},\mathcal{T}_{ansm}\big] &=&n\mathcal{T}_{ansm}  \notag \\
\big[R_{0},\mathcal{T}_{ansm}\big] &=&m\mathcal{T}_{ansm}\   \notag
\end{eqnarray}


\subsection{A KM algebra associated to $\mathbb{S}^3_F$}

Consider now a `softened' three-sphere $\mathbb{S}_{F}^{3}$ with a deformed
metric $g_{MN}^{\mu }$ such that $\sqrt{\det {(g_{MN}^{\mu })}}=F(\theta
,\psi ,\varphi )\sin \theta $. The scalar product on $L^{2}(\mathbb{S}%
_{F}^{3})$ reduces to
\begin{equation*}
(f,g)_{F}=\frac{1}{16\pi ^{2}}\int\limits_{0}^{\pi }\text{d}\theta
\int\limits_{0}^{2\pi }\text{d}\varphi \int\limits_{0}^{4\pi }\text{d}\psi
\;\sin \theta \;F(\theta ,\psi ,\varphi )\;\overline{f(\theta ,\psi ,\varphi
)}g(\theta ,\psi ,\varphi )\ .
\end{equation*}%
The results of Sections \ref{sec:def} and \ref{sec:KKMM} enable us to define KM algebras
associated to $\mathbb{S}_{F}^{3}$:

\begin{enumerate}
\item The Hilbert basis of $L^{2}(\mathbb{S}_{F}^{3})$ is
\begin{equation*}
\mathcal{B}_{F}=\Big\{\psi _{nsm}^{F}(\theta ,\psi ,\varphi
)=\frac{1}{\sqrt{F\left( \theta ,\psi ,\varphi \right) }}\psi _{nsm}\left(
\theta ,\psi ,\varphi \right) =P_{nsm}^{~~~prq}\psi _{prq}(\theta ,\psi ,\varphi )\ ,\ \ s\in \frac{1}{2}%
\mathbb{N},-s\leq n,m\leq s\Big\},
\end{equation*}
where the coefficients $P_{nsm}^{~~~prq}$ are defined according to the
general formula (\ref{eq:P}).

\item The Hermitean operators associated to $\mathbb{S}_{F}^{3}$ are $L_{0}$
and $R_{0}$ themselves.

\item The two-cocycles associated to $L_{0}$ and $R_{0}$ are given by
\begin{eqnarray*}
\omega _{\alpha }^{F}\left( T_{ansm}^{F},T_{bn^{\prime }s^{\prime }m^{\prime
}}^{F}\right)  &=&P_{nsm}^{~~~prq}P_{n^{\prime }s^{\prime }m^{\prime
}}^{~~~~~p^{\prime }r^{\prime }q^{\prime }}\omega _{\alpha }\left(
T_{aprq},T_{bp^{\prime }r^{\prime }q^{\prime }}\right)  \\
&=&\eta _{ab}\left( -1\right) ^{m-n}P_{nsm}^{~~~prq}P_{n^{\prime }s^{\prime
}m^{\prime }}^{~~~~~p^{\prime }r^{\prime }q^{\prime }}\delta _{ss^{\prime
}}\delta _{n,-n^{\prime }}\delta _{m,-m^{\prime }}\left( \delta _{\alpha
L}k_{L}p^{\prime }+\delta _{\alpha R}k_{R}q^{\prime }\right) ,
\end{eqnarray*}
where $\alpha=L$ and $R$, respectively.
\end{enumerate}

Thus, from Sections \ref{sec:def} and \ref{sec:KKMM} the KM algebra associated to the `softened'
three-sphere $\mathbb{S}_{F}^{3}$ is
\begin{equation*}
\widetilde{\mathfrak{g}}(\mathbb{S}_{F}^{3})=\Big\{\mathcal{T}%
_{ansm}^{F},L_{0},R_{0},k_{L},k_{R}\ ,\ \ a=1,\cdots ,d,s\in \frac{1}{2}%
\mathbb{N},-s\leq n,m\leq s\Big\}
\end{equation*}%
Taking into account that
\begin{equation*}
\frac{1}{\sqrt{F(\theta ,\psi ,\varphi )}}=\sum_{s\in \frac{1}{2}\mathbb{N}%
}\sum\limits_{n=-s}^{s}\sum\limits_{m=-s}^{s}F^{nsm}\psi _{nsm}(\theta ,\psi
,\varphi ),
\end{equation*}%
the Lie brackets reduce to
\begin{eqnarray}
\big[\mathcal{T}_{ansm}^{F},\mathcal{T}_{a^{\prime }n^{\prime }s^{\prime
}m^{\prime }}^{F}\big] &=&if_{aa^{\prime }}{}^{a^{\prime \prime
}}c_{ss^{\prime }nn^{\prime }mm^{\prime }}^{Fs^{\prime \prime }}\mathcal{T}%
_{a^{\prime \prime }n+n^{\prime }s^{\prime \prime }m+m^{\prime }}^{F}  \notag
\\
&&+\eta _{ab}\left( -1\right) ^{p-q}P_{nsm}^{~~~prq}P_{n^{\prime }s^{\prime
}m^{\prime }}^{~~~p^{\prime }r^{\prime }q^{\prime }}\delta _{rr^{\prime
}}\delta _{p,-p^{\prime }}\delta _{q,-q^{\prime }}\left( k_{L}p^{\prime
}+k_{R}q^{\prime }\right) ; \\
\big[L_{0},\mathcal{T}_{ansm}^{F}\big] &=&pP_{nsm}^{~~~prq}\mathcal{T}_{aprq}^{F};  \notag \\
\big[R_{0},\mathcal{T}_{ansm}^{F}\big] &=&qP_{nsm}^{~~~prq}\mathcal{T}_{aprq}^{F} .  \notag
\end{eqnarray}%
where the `Clebsch-Gordan coefficients' of $\mathbb{S}_{F}^{3}$
 $c_{ss^{\prime }nn^{\prime }mm^{\prime }}^{Fs^{\prime \prime }}$ are
defined by an expression analogous to \eqref{eq:Cmu}.


\subsubsection{$\widetilde{\mathbb{S}}^3$ and its physical applications}


An important example of `softened' three-sphere is the \textit{Berger
three-sphere}, also named squashed three-sphere, denoted by $\widetilde{%
\mathbb{S}}^{3}$. Introducing the right-invariant one-forms \eqref{eq:RI1F}
and following Berger \cite{Berger}, the (doubly) squashed three-sphere $%
\widetilde{\mathbb{S}}^{3}$ is endowed with the following metric:
\begin{eqnarray}
\text{d}s^{2} &=&\omega _{1}^{2}+b^{2}\omega _{2}^{2}+c^{2}\omega _{3}^{2}
\notag  \label{eq:gSS3} \\
&=&\Big(\sin ^{2}\psi +b^{2}\cos ^{2}\psi \Big)\text{d}\theta ^{2}+c^{2}%
\text{d}\psi ^{2}+\Big(\sin ^{2}\theta \cos ^{2}\psi +b^{2}\sin ^{2}\psi
\sin ^{2}\theta +c^{2}\cos ^{2}\theta \Big)\text{d}\varphi ^{2}  \notag \\
&&+2(b^{2}-1)\sin \psi \cos \psi \sin \theta \text{d}\theta \text{d}\varphi
+2c^{2}\cos ^{2}\theta \text{d}\varphi \text{d}\psi \ ,
\end{eqnarray}%
where $b,c>0$ are named \textit{squashing parameters}. Thus, the metric
tensor takes the form
\begin{equation*}
g=%
\begin{pmatrix}
\sin ^{2}\psi +b^{2}\cos ^{2}\psi & 0 & (1-b^{2})\sin \psi \cos \psi \sin
\theta \\
0 & {c}^{2} & c^{2}\cos \theta \\
(b^{2}-1-1)\sin \psi \cos \psi \sin \theta & {c}^{2}\cos \theta & \sin
^{2}\theta \cos ^{2}\psi +b^{2}\sin ^{2}\psi \sin ^{2}\theta +c^{2}\cos
^{2}\theta%
\end{pmatrix}%
\end{equation*}%
and
\begin{equation*}
\sqrt{\det g}=bc\sin \theta \ .
\end{equation*}%
As $\omega _{1},\omega _{2},\omega _{3}$ in \eqref{eq:RI1F} are
right-invariant (but not left-invariant) one-forms, the metric %
\eqref{eq:gSS3} is invariant under the right-action of SU$(2)$, but not
under the left-action of SU$(2)$. Thus, $R_{\pm },R_{0}$ generate isometries
of the squashed sphere. Moreover, the scalar curvature of $\widetilde{%
\mathbb{S}}^{3}$ is given by
\begin{equation*}
R=-\frac{b^{4}+(c^{2}-1)^{2}-2b^{2}(c^{2}+1)}{2b^{2}c^{2}},
\end{equation*}%
which for suitable choices of $b$ and $c$, can be made
negative; for instance, by setting $b=1$, one obtains%
\begin{equation}
R=2-c^{2}/2<0\Leftrightarrow c^{2}>4\ .  \notag
\end{equation}

The squashed three-sphere $\widetilde{\mathbb{S}}^{3}$ with two squashing
parameters $b$ and $c$ provides a rich framework for understanding various
physical phenomena. Indeed, $b$ and $c$ introduce anisotropic scaling in the
directions corresponding to the $SU(2)$ right-invariant one-forms $\omega
_{2}$ and $\omega _{3}$, and therefore the original $SO(4)$ isometry of $%
\mathbb{S}^{3}$ gets reduced, e.g. typically to $SU(2)$, for generic values
of $b$ and $c$.

In the mathematics context, $\widetilde{\mathbb{S}}^{3}$ was considered by
Hitchin \cite{Hitchin} in his discussion of the space of harmonic spinors
(i.e., the null space of the Dirac operator) on a manifold: in fact, $%
\widetilde{\mathbb{S}}^{3}$ is a notable illustration of the fact that the
number of harmonic spinors is not a topological invariant of the manifold,
but rather it depends on the particular metric, as well. In physics, scalar
quantum field theory on $\widetilde{\mathbb{S}}^{3}$ has been investigated
by several authors \cite{sqft1,sqft2,sqft3}, mainly due to its appearance as
(a particular case of) the spatial section of the mixmaster cosmological
model \cite{mixmaster}.

Moreover, $\widetilde{\mathbb{S}}^{3}$ naturally arises in flux
compactifications and Kaluza-Klein (KK) reductions. For example,
compactifications of 11-dimensional supergravity or 10-dimensional string
theory on $\widetilde{\mathbb{S}}^{3}$ lead to modifications in the
low-energy effective theory, where the parameters $b$ and $c$ break the
internal symmetry, affecting the spectrum of KK modes \cite{KK1,KK2,KK3}. As
first discussed in \cite{gibbons1979}, these deformations can break part of
the supersymmetry and affect the vacuum structure, the cosmological constant
and the gauge couplings in the lower-dimensional theory. Within the AdS/CFT
correspondence, $\widetilde{\mathbb{S}}^{3}$ plays a significant role, for
instance in the \textit{AdS$_{3}$} solutions of supergravity and string
theory, since the geometry of the internal manifold, controlled by the
squashing parameters, affects the dual two-dimensional conformal field
theory (CFT). On the other hand, within the gauge/gravity duality, the
gravity duals of supersymmetric gauge theories on $\widetilde{\mathbb{S}}%
^{3} $ (with various types of squashing) have been investigated \cite{M1,M2}%
, as well. Compactifications involving $\widetilde{\mathbb{S}}^{3}$ are also
important in M-theory, because $\widetilde{\mathbb{S}}^{3}$ can form part of
the internal spaces with $\mathit{G}_{2}$ holonomy, as seen in supergravity
solutions with fluxes. It is also worth recalling here that the effective
actions for both scalars and fermions on $\widetilde{\mathbb{S}}^{3}$ have
been obtained in \cite{D1,D2}, with the aim to compare with the AdS/CFT
results of \cite{AdS1,AdS2,AdS3}; notice that the relation is far from being
obvious, since the regimes where the results are expected to apply are very
different (cf. e.g. \cite{TR} for a discussion).

We should recall that $\widetilde{\mathbb{S}}^{3}$ is also a key tool in
localization techniques within the aforementioned supersymmetric gauge
theories; specifically, in three-dimensional $\mathcal{N}=2$ supersymmetric
gauge theories, placing the theory on $\widetilde{\mathbb{S}}^{3}$ allows
for the exact computation of partition functions and other observables
through localization. Hama, Hosomichi, and Lee \cite{hama2011} showed that
the squashing parameters $b$ and $c$ modify the background geometry,
influencing the preserved supersymmetry. This affects the localization
\textit{locus} of the path integral, which in turn modifies the resulting
physical observables, such as the exact partition function and the Wilson
loops. These computations are particularly useful for studying
non-perturbative effects and for testing dualities, such as mirror symmetry
and (three-dimensional) Seiberg-like dualities. In general, $b$ and $c$
induce a continuous deformation of the background geometry, thus enabling
the investigation of different phases of the theory. For instance, in
supersymmetric gauge theories with exact localization, the partition
function is an integral over the moduli space of flat connections, and the
squashing parameters alter the effective action and measure of the integral,
as noted again in \cite{hama2011}.

The invariant measure on the squashed sphere reads
\begin{eqnarray}
\int \limits_{{\mathbb{S}^3}} \text{d} \mu(\widetilde{\mathbb{S}^3})=\frac 1
{16 \pi^2}\int \limits_{{\mathbb{S}^3}} \sqrt{\det g} \;\text{d} \theta \;%
\text{d} \varphi\; \text{d}\psi =\frac 1 {16 \pi^2} \int \limits_{0}^{\pi}
cb \sin \theta \;\text{d} \theta \int \limits_{0}^{2\pi} \text{d} \varphi
\int \limits_{0}^{4\pi} \text{d} \psi \ .
\end{eqnarray}
This means that
\begin{eqnarray}
\mathcal{B}_{\widetilde{\mathbb{S}}^3}=\Big\{\widetilde{\psi}_{nsm}= \frac 1
{\sqrt{bc}} \psi_{nsm}\ , \ \ s \in \frac 12 \mathbb{N}\ , \ \ -s \le n,m\le
s\Big\}  \notag
\end{eqnarray}
with $\psi_{nsm}$ corresponding to the matrix elements of SU$(2)$ (see e.g. %
\eqref{eq:HS3}) constitutes a Hilbert basis of the squashed sphere:
\begin{eqnarray}
(\widetilde{\psi}_{nsm},\widetilde{\psi}_{n^{\prime }s^{\prime }m^{\prime
}})= \int \limits_{{\mathbb{S}^3}} \text{d} \mu(\widetilde{\mathbb{S}^3})\;
\overline{\widetilde{\psi}_{nsm}}(\theta,\varphi,\psi)\;\widetilde{\psi}%
_{n^{\prime }s^{\prime }m^{\prime }}(\theta,\varphi,\psi)
=\delta_{ss^{\prime }} \delta_{nn^{\prime }} \delta_{mm^{\prime }} \ .
\notag
\end{eqnarray}
Note that, as mentioned in Section \ref{sec:KMG}, $\widetilde{\psi}_{nsm}$
are in representation of the left and right actions of the group SU$(2)$,
but only the right action is an isometry of the squashed sphere.

From Section \ref{sec:CC} the product $\widetilde{\psi}_{nsm}(\theta,\psi,%
\phi)\widetilde{\psi}_{n^{\prime }s^{\prime }m^{\prime }}(\theta,\psi,\phi)$
is straightforward. The KM algebra associated to the squashed sphere $%
\widetilde{\mathfrak{g}}(\widetilde{\mathbb{S}}^3)$ follows directly, and it
is isomorphic to the KM algebra associated to the usual three-sphere. Since
the usual sphere is obtained taking the limit $b,c\to 1$, considering this
limit we recover an isomorphic realization of the KM algebra of $\mathbb{S}%
^3 $ from the KM algebra of the squashed sphere.


\section{`Softening' of $\mathbb{S}^{2}$ and the associated KM algebra}

As a third example, we consider now the KM algebra associated to the
`softening'of the two-sphere $\mathbb{S}^{2}$.

\subsection{The two-sphere $\mathbb{S}^2$}

\label{sec:S2} The two-sphere $\mathbb{S}^{2}$ is given by the symmetric
space $SO(3)/SO(2)\simeq SU(2)/U(1)=\mathbb{C}P^{1}$, namely by the complex
projective line (here $\simeq $ denotes isomorphism of homogeneous spaces).
The \textit{complex projective line} $\mathbb{C}P^{1}$ has profound
applications in several areas of theoretical physics, particularly in gauge
theory, string theory, and twistor theory, because the isomorphism $\mathbb{C%
}P^{1}\simeq \mathbb{S}^{2}$ makes it a useful model for understanding
internal symmetries and topological properties in physical systems. In gauge
theory, for instance, $\mathbb{C}P^{1}$ plays a crucial role in the study of
monopole solutions. The celebrated 't Hooft-Polyakov monopole solution,
which arises in non-Abelian gauge theory, uses $\mathbb{C}P^{1}$ as the
internal symmetry space that describes the direction of the Higgs field at
spatial infinity. The mapping from spatial infinity $\mathbb{S}^{2}$ to $%
\mathbb{C}P^{1}$, classified by the homotopy group $\pi
_{2}(SU(2)/U(1))\cong \mathbb{Z}$, gives rise to the topological charge of
the monopole, which corresponds to the magnetic charge \cite%
{tHooft1974,Polyakov1974}.

In string theory, $\mathbb{CP}^{1}$ arises naturally in sigma models, where
it serves as the target space of two-dimensional field theories describing
strings. On the other hand, in topological string theory $\mathbb{C}P^{1}$
provides a simple setting for computing Gromov-Witten invariants, counting
the number of holomorphic maps from the string worldsheet to the target
space (see e.g. \cite{MirrorSymmetry1997}). Additionally, $\mathbb{C}P^{1}$
often appears as part of the internal geometry in compactifications of
higher-dimensional theories, such as compactifications of type II string
theory on Calabi-Yau manifolds, where $\mathbb{CP}^{1}$ can represent
two-cycles within the compactified space, in turn determining the low-energy
effective theory and the spectrum of BPS states.

Furthermore, in twistor theory $\mathbb{C}P^{1}$ appears in the context of
describing spacetime in terms of complex geometry. Indeed, the twistor space
$\mathbb{C}P^{3}$, which encodes information about the four-dimensional
Minkowski spacetime, is fibered over $\mathbb{CP}^{1}$, with each point in
spacetime corresponding to a projective line in twistor space. This
reformulation of spacetime, where points are replaced by lines in a complex
projective space, allows the description of gravitational and gauge field
theories in terms of holomorphic structures. The incidence relation between
points in twistor space and lines in $\mathbb{C}P^{1}$ is given by a simple
geometric condition, leading to the so-called Penrose transform, converting
solutions of wave equations in spacetime into holomorphic functions in
twistor space \cite{Penrose1986}. This geometric framework has been
successfully applied to study the scattering amplitudes of gauge theories
and gravity, providing new insights into the structure of field theories.

In the following, we assume that $Q \in U(1) \subset SU(2)$ is given by
\begin{eqnarray}
Q= e^{ i\theta R_0}\ .  \notag
\end{eqnarray}
The points $(\theta,\psi, \varphi = \text{cons.})$ parameterise points on
the manifold $\mathbb{S}^2 \cong SU(2)/U(1) \subset \mathbb{S}^3 \cong
SU(2). $ With this parameterisation, for the level surfaces $\varphi=$ \text{%
cons.} we have, on the one hand
\begin{eqnarray}
L_\pm= e^{\pm i\psi} \Big(i\cot \theta \partial_\psi \pm \partial_\theta\Big)%
\ , \ \ L_0= - i \partial_\psi \ ,  \notag
\end{eqnarray}
as well as the relation
\begin{eqnarray}
(f,g) &=& \frac 1{16 \pi^2}\int \limits_{0}^{\pi} \sin\theta \text{d} \theta
\int\limits_0^{2 \pi} \text{d} \varphi \int\limits_0^{4 \pi} \text{d} \psi\;
\overline{f(\theta,\psi,\varphi)} \; {g(\theta,\psi,\varphi)}  \notag \\
&=&\Big|_{\varphi = \text{cons.}}\; \frac1{8\pi} \int \limits_{0}^{\pi} \sin
\theta \;\text{d} \theta \int\limits_0^{4 \pi} \text{d} \psi \overline{f
\theta,\psi, \varphi =\text{cons.})} \; {g( \theta,\psi, \varphi =\text{cons.%
})} \ .  \notag
\end{eqnarray}
The overall factor is due to the fact that we cover twice the two-sphere.
From now on we set $\psi\in[0,2\pi[$ and substitute the normalization factor
in the integral above by $1/4\pi$. Since the harmonic analysis on the SU$%
(2)/ $U$(1)$ coset (and the related matrix elements) must be U$(1)-$%
invariant, the only matrix elements which appear are those which are $U(1)-$%
invariant, thus corresponding to the matrix elements with $m=0$ (bosonic
case with $s \in \mathbb{N}$), which in turn reduce to spherical harmonics.
We thus have the explicit polynomial expression for spherical harmonics (the
sign $(-1)^ {s}$ below is to agree with standard definition of spherical
harmonics):
\begin{eqnarray}  \label{eq:Y}
\begin{array}{c}
Y_{sn}(\theta,\psi) =\frac{(-1)^{{s}+\frac{|n|+n}2}\sqrt{2s+1} }{|n|!} \sqrt{
\frac{(s+|n|)!}{(s-|n|)!}} e^{ i n \psi} \cos^{|n|} \frac \theta 2
\sin^{|n|} \frac \theta 2 {}_2F_1(|n|-s,|n|+s+1;1+|n|; \; \sin^2 \frac
\theta 2)%
\end{array}%
\end{eqnarray}
Thus $\{Y_{sn}, s \in \mathbb{N}, -s\le n\le s\}$ constitutes a Hilbert
basis of $L^2(\mathbb{S}^2)$ and we have
\begin{eqnarray}
(Y_{sn},Y_{s^{\prime }n^{\prime }})=\delta_{ss^{\prime }} \delta_{nn^{\prime
}} \ .  \notag
\end{eqnarray}
As in Section \ref{sec:CC}, we also have
\begin{eqnarray}  \label{eq:YY}
Y_{s_1n_1}(\theta,\psi) Y_{s_2n_2}(\theta,\psi) = \sum
\limits_{S=|s_1-s_2|}^{s_1+s_2} c_{s_1 s_2}^{S}{}_{n_1n_2}\; Y_{S
n_1+n_2}(\theta,\psi)
\end{eqnarray}
with
\begin{eqnarray}  \label{eq:CCSO3}
c_{s_1 s_2}^{S}{}_{n_1n_2 } \equiv \sqrt{ \frac{(2s_1+1)(2s_2+1)}{2S+1}}
\begin{pmatrix}
s_1 & s_2 & S \\
n_1 & n_2 & n_1+n_2%
\end{pmatrix}%
\;
\begin{pmatrix}
s_1 & s_2 & S \\
0 & 0 & 0%
\end{pmatrix}%
\end{eqnarray}


\subsection{KM and Virasoro algebras associated to $\mathbb{S}^2$}


The KM algebra associated to $\mathbb{S}^2$ follows directly from %
\eqref{eq:CCSO3} and the results in \cite{mrm}. Indeed, we have
\begin{eqnarray}
\widetilde{\mathfrak{g}}(\mathbb{S}^2)= \Big\{\mathcal{T}_{a\ell m}, J_0, k\
, a=1,\cdots \dim\mathfrak{g}\ , \ell \in \mathbb{N}\ , -\ell\le m \le \ell %
\Big\} \ ,  \notag
\end{eqnarray}
with the central charge $k$ associated to the two-cocycle:
\begin{eqnarray}
\omega(x,y) = \frac k{4\pi} \int \limits_0^\pi \sin \theta\; \text{d}\theta
\int \limits_0^{2\pi} \text{d}\psi \;\big<x, J_0 y\big>_0 \ .  \notag
\end{eqnarray}
The Lie brackets take the form (see \cite{mrm})
\begin{eqnarray}  \label{eq:KM-S2}
\big[\mathcal{T}_{a_1\ell_1 m_1},\mathcal{T}_{a_2\ell_2 m_2}\big]&=& i
f_{a_1 a_2}{}^{a_ 3} c_{\ell_1,m_1,\ell_2,m_2}^{\ell_3} \mathcal{T}%
_{a_3\ell_3 m_1+m_2} + (-1)^{m_1} k m_2\; \eta_{a_1 a_2}
\delta_{\ell_1,\ell_2}\delta_{m_1,- m_2} \ , \\
\big[J_0,\mathcal{T}_{a \ell m}\big]&=& m\mathcal{T}_{a \ell m} \ .  \notag
\end{eqnarray}

An analogue of the Virasoro algebra on the two-sphere was defined in \cite%
{rm}. We briefly reproduce here the results in a way suitable to extend this
algebra on the `softened' sphere. To proceed, set
\begin{equation*}
Y_{\ell m}(\theta ,\psi )=Q_{\ell m}(\theta )e^{im\psi }\ .
\end{equation*}%
Note also that for $f\in L^{2}(\mathbb{S}^{2})$ we have
\begin{equation*}
f(\theta ,\psi )=\sum\limits_{\ell =0}^{+\infty }\sum\limits_{m=-\ell
}^{\ell }f^{\ell m}Y_{\ell m}(\theta ,\psi )=\sum\limits_{m\in \mathbb{Z}}%
\Big(\sum\limits_{\ell \geq |m|}f^{\ell m}Q_{\ell m}(\theta )\Big)e^{im\psi
}\ .
\end{equation*}%
The second summation is not usual, but more appropriate on purpose. The
orthogonality between spherical harmonics implies
\begin{equation*}
\frac{1}{2}\int\limits_{0}^{2\pi }\sin \theta \;\text{d}\theta \;Q_{\ell
m}(\theta )Q_{\ell ^{\prime }m}(\theta )=\delta _{\ell \ell ^{\prime }}
\end{equation*}%
and \eqref{eq:YY} leads to
\begin{equation*}
Q_{\ell m}(\theta )Q_{\ell ^{\prime }m^{\prime }}(\theta )=c_{\ell ,m,\ell
^{\prime },m^{\prime }}^{\ell ^{\prime \prime }}Q_{\ell ^{\prime \prime
},m+m^{\prime }}(\theta )\ .
\end{equation*}

We now introduce
\begin{eqnarray}
\ell_{\ell m} = i Q_{\ell m}(\theta) e^{im\psi} \partial_\psi \ .
\end{eqnarray}
The set $\{\ell_{\ell m}, m \in \mathbb{N}, \ell\ge |m|\}$ constitutes a
subset of the vector fields on the two-sphere with Lie brackets:
\begin{eqnarray}  \label{eq:wittS2}
\big[\ell_{\ell m}, \ell_{\ell^{\prime }m^{\prime }}\big] = (m-m^{\prime })
c_{\ell, m,\ell^{\prime },m^{\prime }}^{\ell^{\prime \prime }}
\ell_{\ell^{\prime },m+m^{\prime }} \ .
\end{eqnarray}
This algebra admits a non-trivial two-cocycle, which is analogous to the Gel'fand--Fuks cocycle for the
Virasoro algebra:
\begin{eqnarray}
\omega(f,g) = -\frac i{12} \frac c{4\pi} \int \limits_0^\pi \sin\theta \text{%
d}\theta \int \limits_0^{2\pi} \text{d}\psi f(\theta,\psi) \partial_\psi^3
g(\theta,\psi) \ .  \notag
\end{eqnarray}
Evaluated on the vector fields \eqref{eq:wittS2}, this gives
\begin{eqnarray}  \label{eq:cocy}
\omega(\ell_{\ell,m},\ell_{\ell^{\prime },m^{\prime }}) = \frac c{12} m^3
\delta_{\ell\; \ell^{\prime }} \delta_{m,-m^{\prime }} \ .
\end{eqnarray}
Let $\mathcal{L}_{\ell m}$ be the generators of the centrally extended
algebra \eqref{eq:wittS2} by the cocycle \eqref{eq:cocy}. Define now $%
L_{\ell m} = \mathcal{L}_{\ell m} + \frac c{24} \delta_{m,0}$. The Virasoro
algebra of the two-sphere
\begin{eqnarray}
\mathrm{Vir}(\mathbb{S}^2) = \Big\{L_{\ell,m}\ ,\ \ m \in \mathbb{Z}\ ,\ \
\ell \ge |m|\ ,\ \ c \Bigg\}  \notag
\end{eqnarray}
has Lie brackets
\begin{eqnarray}  \label{eq:VirS2}
\big[L_{\ell m}, L_{\ell^{\prime }m^{\prime }}\big]= (m-m^{\prime })\;
c_{\ell,m,\ell^{\prime },m^{\prime }}^{\ell^{\prime \prime }}
L_{\ell^{\prime \prime }m+m^{\prime }} +\frac c{12} (m^3-m)\delta_{\ell
\ell^{\prime }} \delta_{m,-m^{\prime }} \ .
\end{eqnarray}
We observe that this algebra was defined in \cite{rm} in a slightly
different way.

The KM and Virasoro algebra of the two-sphere admits a semi-direct structure
$\widetilde{\mathfrak{g}}(\mathbb{S}^2)\rtimes$Vir$(\mathbb{S}^2)$ with
action of $L_{\ell m}$ on $\mathcal{T}_{a \ell^{\prime }m^{\prime }}$:
\begin{eqnarray}  \label{eq:VKM}
\big[L_{\ell m},\mathcal{T}_{a \ell^{\prime }m^{\prime }}\big]=-m^{\prime
}c_{\ell,m,\ell^{\prime },m^{\prime }}^{\ell^{\prime \prime }}\mathcal{T}_{a
\ell^{\prime \prime }m+m^{\prime }} \ ,
\end{eqnarray}
and we have $J_0=-L_{00}$. The algebra $\widetilde{\mathfrak{g}}(\mathbb{S}%
^2)\rtimes$Vir$(\mathbb{S}^2)$ is thus defined by \eqref{eq:KM-S2} and %
\eqref{eq:VirS2} and \eqref{eq:VKM}.

\subsection{A KM algebra associated to $\mathbb{S}^2_F$}

\label{sec:S2d} Consider now a `softened' two-sphere $\mathbb{S}_{F}^{2}$
with a deformed metric $g_{MN}^{\mu }$ such that $\sqrt{\det {(g_{MN}^{\mu })%
}}=F(\theta ,\psi )\sin \theta $. The scalar product on $L^{2}(\mathbb{S}%
_{F}^{2})$ reduces to
\begin{equation*}
(f,g)_{F}=\frac{1}{4\pi ^{2}}\int\limits_{0}^{\pi }\text{d}\theta
\int\limits_{0}^{2\pi }\text{d}\psi \sin \theta \;F(\theta ,\psi )\;%
\overline{f(\theta ,\psi )}g(\theta ,\psi )\ .
\end{equation*}%
The results of Sections \ref{sec:def} and \ref{sec:KKMM} enable us to define KM algebras
associated to $\mathbb{S}_{F}^{2}$:

\begin{enumerate}
\item The Hilbert basis of $L^{2}(\mathbb{S}_{F}^{2})$ is
\begin{equation*}
\mathcal{B}_{F}=\Big\{Y_{\ell m}^{F}(\theta ,\psi )=\frac{1}{\sqrt{F\left( \theta ,\psi \right) }}Y_{\ell m}\left( \theta
,\psi \right) =P_{\ell m}^{~~\ell
^{\prime }m^{\prime }}Y_{\ell ^{\prime }m^{\prime }}(\theta ,\psi )\ ,\ \
\ell \in \mathbb{N},-\ell \leq m\leq \ell \Big\},
\end{equation*}
where the coefficients $P_{\ell m}^{~~\ell ^{\prime }m^{\prime }}$ are
defined according to the general formula (\ref{eq:P}).

\item the Hermitean operator associated to $\mathbb{S}_{F}^{2}$ is $J_{0}$
itself.

\item the two-cocycle associated to $J_{0}$ is given by
\begin{eqnarray*}
\omega ^{F}\left( T_{a_{1}\ell _{1}m_{1}}^{F},T_{a_{2}\ell
_{2}m_{2}}^{F}\right)  &=&P_{\ell _{1}m_{1}}^{~~~\ell _{1}^{\prime
}m_{1}^{\prime }}P_{\ell _{2}m_{2}}^{~~~~~\ell _{2}^{\prime }m_{2}^{\prime
}}\omega \left( T_{a_{1}\ell _{1}^{\prime }m_{1}^{\prime }},T_{a_{2}\ell
_{2}^{\prime }m_{2}^{\prime }}\right)  \\
&=&k\eta _{a_{1}a_{2}}\left( -1\right) ^{m_{1}^{\prime }}m_{2}^{\prime
}P_{\ell _{1}m_{1}}^{~~~\ell _{1}^{\prime }m_{1}^{\prime }}P_{\ell
_{2}m_{2}}^{~~~~~\ell _{2}^{\prime }m_{2}^{\prime }}\delta _{\ell
_{1}^{\prime }\ell _{2}^{\prime }}\delta _{m_{1}^{\prime },-m_{2}^{\prime }}.
\end{eqnarray*}
\end{enumerate}

Thus from Sections \ref{sec:def} and \ref{sec:KKMM}, we define the analogue of the loop algebra $%
\mathfrak{g}(\mathbb{S}_{F}^{2})=\big\{T_{a\ell m}=T_{a}P_{\ell m}^{~~\ell
^{\prime }m^{\prime }}Y_{\ell ^{\prime }m^{\prime }}(\theta ,\psi )\big\}$ and the corresponding centrally extended Lie algebra $\widetilde{\mathfrak{g}}(\mathbb{S}_{F}^{2})$,
\begin{equation*}
\widetilde{\mathfrak{g}}(\mathbb{S}_{F}^{2})=\Big\{\mathcal{T}_{a\ell
m}^{F},J_{0},k\ ,\ \ a=1,\cdots ,d,\ell \in \mathbb{N},-\ell \leq m\leq \ell %
\Big\}.
\end{equation*}%
Using the relation
\begin{equation*}
\frac{1}{\sqrt{F(\theta ,\psi )}}=\sum_{\ell \in \mathbb{N}%
}\sum\limits_{m=-\ell }^{\ell }F^{\ell m}Y_{\ell m}(\theta ,\psi )
\end{equation*}%
the Lie brackets reduce to
\begin{eqnarray}
\big[\mathcal{T}_{a_{1}\ell _{1}m_{1}}^{F},\mathcal{T}_{a_{2}\ell
_{2}m_{2}}^{F}\big] &=&if_{a_{1}a_{2}}{}^{a_{3}}c_{\ell _{1},m_{1},\ell
_{2},m_{2}}^{F\ell }\mathcal{T}_{a_{3},\ell
_{3},m_{1}+m_{2}}^{F}+(-1)^{m_{1}^{\prime }}km_{2}^{\prime }\eta
_{a_{1}a_{2}}P_{\ell _{1}m_{1}}^{~~\ell _{1}^{\prime }m_{1}^{\prime
}}P_{\ell _{2}m_{2}}^{~~\ell _{2}^{\prime }m_{2}^{\prime }}\delta _{\ell
_{1}^{\prime },\ell _{2}^{\prime }}\delta _{m_{1}^{\prime },-m_{2}^{\prime
}},   \notag \\
\big[J_{0},\mathcal{T}_{a\ell m}^{F}\big] &=&m^{\prime }P_{\ell m}^{~~\ell ^{\prime }m^{\prime }}\mathcal{T}_{a\ell ^{\prime
}m^{\prime }}^{F} \label{eq:KM-S2F}.
\end{eqnarray}%
where the `Clebsch-Gordan coefficients' of $\mathbb{S}_{F}^{2}$ $c_{\ell m\ell ^{\prime }m^{\prime }}^{F\ell ^{\prime \prime }}$ are
defined by an expression analogous to \eqref{eq:Cmu}.\newline

If we now define an analogue of Virasoro algebra on the `softened'
two-sphere $\mathbb{S}_{F}^{2}$, by arguments similar than in Section \ref%
{sec:defS1}, one can show that the Virasoro algebra of the `softened'
two-sphere is isomorphic to the Virasoro algebra on the two-sphere.

\subsubsection{$\widetilde{\mathbb{S}}^2$ and its physical applications}

The \textit{squashed two-sphere} $\mathbb{S}_{b}^{2}\equiv \mathbb{\tilde{S}}%
^{2}$ is a `softened' version of the usual two-sphere $\mathbb{S}^{2}$,
where the geometry is compressed or stretched along one axis, breaking the
spherical symmetry while maintaining some degree of residual symmetry,
typically $U(1)$. The metric of the squashed two-sphere is often written as

\begin{equation*}
ds^{2}=d\theta ^{2}+b^{2}\sin ^{2}\theta \,d\phi ^{2}
\end{equation*}%
where $b$ is the squashing parameter, $\theta \in \lbrack 0,\pi ]$ and $\phi
\in \lbrack 0,2\pi )$ are the usual spherical coordinates. For $b=1$, this
reduces to the standard metric on $\mathbb{S}^{2}$, but when $b\neq 1$, the
symmetry is reduced to $U(1)$, representing squashing along one of the axes.
This squashed geometry has important applications in theoretical physics,
especially in gauge theory \cite{Benini2012}, string theory \cite{Witten1982}%
, and supersymmetric field theory \cite{Martelli2006}.

In gauge theory, $\mathbb{\tilde{S}}^{2}$ plays a significant role in the
localization of supersymmetric field theories. The exact partition function
of two-dimensional $\mathcal{N}=2$ supersymmetric gauge theories on $\mathbb{%
\tilde{S}}^{2}$ can be computed using localization techniques. In this
context, the metric deformation encoded in the squashing parameter $b$
modifies the background geometry, which affects the supersymmetry-preserving
configurations and the resulting partition function. The action for a
supersymmetric field theory placed on $\mathbb{\tilde{S}}^{2}$ is deformed
by the squashing, and the partition function can be expressed as an integral
over the Coulomb branch, depending on the squashing parameter \cite%
{Benini2012}. The exact partition function is written as $Z(b)=\int d\sigma
\,e^{-S_{\text{eff}}(\sigma ,b)}$, where $\sigma $ denotes the scalar field
in the vector multiplet, and $S_{\text{eff}}$ is the effective action that
depends on the squashing parameter $b$.

In string theory, $\mathbb{\tilde{S}}^{2}$ appears in the context of flux
compactifications and as an internal space in string sigma models. For
instance, in compactifications of string theory on non-trivial backgrounds, $%
\mathbb{\tilde{S}}^{2}$ provides a natural compactification space with
reduced symmetry that still preserves some supersymmetry. This
compactification can lead to interesting low-energy effective theories where
the squashing parameter $b$ controls the amount of symmetry breaking.
Additionally, in string sigma models $\mathbb{\tilde{S}}^{2}$ provides a
target space for two-dimensional field theories that describe strings
propagating on deformed geometries. The deformation of the target space
modifies the spectrum of the theory, leading to shifts in the masses of
excitations and affecting the dynamics of the system \cite{Witten1982}.

\textit{Last but not least}, within the AdS/CFT correspondence, as mentioned
above squashed spheres, thus including $\mathbb{\tilde{S}}^{2}$, arise in
the context of holographic dualities. In particular, squashed spheres appear
as internal compactification spaces in AdS spacetimes, where the dual field
theory resides on the boundary of the AdS space. The squashing of the
internal sphere leads to deformations of the dual field theory, affecting
its operator spectrum and the correlation functions. For example, $\mathbb{%
\tilde{S}}^{2}$ can modify the dual conformal field theory by breaking
certain symmetries while preserving others, thereby providing a useful tool
to study symmetry-breaking phenomena in the holographic setting \cite%
{Martelli2006}.

The invariant measure on the squashed sphere reads
\begin{eqnarray}
\int \limits_{{\mathbb{S}^2}} \text{d} \mu(\widetilde{\mathbb{S}^2}) =\frac
b {4 \pi} \int \limits_{0}^{\pi} \sin \theta \text{d} \theta \int
\limits_{0}^{2\pi} \text{d} \psi \ .
\end{eqnarray}
This means that
\begin{eqnarray}
\mathcal{B}_{\widetilde{\mathbb{S}}^2}=\Big\{\widetilde{Y}_{\ell m}= \frac 1
{\sqrt{b}} Y_{\ell m}\ , \ \ \ell \in \mathbb{N}\ , \ \ -\ell \le m\le \ell%
\Big\}  \notag
\end{eqnarray}
with $Y_{\ell m}$ corresponding to the matrix elements of $\mathbb{S}^2$
(see e.g. \eqref{eq:Y}) constitutes a Hilbert basis of the squashed sphere:
\begin{eqnarray}
(\widetilde{Y}_{\ell m},\widetilde{Y}_{\ell^{\prime }m^{\prime }})
=\delta_{\ell \ell^{\prime }} \delta_{mm^{\prime }} \ .  \notag
\end{eqnarray}
Note that, as mentioned in Section \ref{sec:KMG}, $\widetilde{Y}_{\ell m}$
are in a representation of SO$(3)$, but the isometry of the squashed sphere
reduces to U$(1)$. From Section \ref{sec:S2}, the description of the product
$\widetilde{Y}_{\ell m}(\theta,\psi)\widetilde{Y}_{\ell^{\prime }m^{\prime
}}(\theta,\psi)$ is straightforward. The KM algebra associated to the
squashed sphere $\widetilde{\mathfrak{g}}(\widetilde{\mathbb{S}}^2)$ follows
directly, and it is isomorphic to the KM
algebra associated to the two-sphere. Since the usual sphere is obtained
taking the limit $b\to 1$, considering this limit we recover an isomorphic
realization of the KM algebra of $\mathbb{S}^2$ from the KM algebra of the
squashed sphere.

\section{Conclusions}

In this paper, we have proposed another generalization of
infinite-dimensional (and infinite-rank) generalized KM algebras $\mathfrak{g}\left( \mathcal{M}\right) $ (and their central extensions $\mathfrak{\tilde{g}}\left(\mathcal{M}\right)$) introduce
in \cite{mrm,rmm2}. Namely, we have set $\mathcal{M}=G_{c}$, and considered the compact group manifold $%
G_{c} $ to be `deformed' into a so-called `soft' group manifold $G_{c}^{\mu
} $, locally diffeomorphic to $G_{c}$ itself. This `softening' procedure
lies at the core of the group-geometric approach to (super)gravity and
superstring theories, and it was extensively studied from the 1970s onwards
by Tullio Regge and his research group, being further developed later by
Riccardo D'Auria and Pietro Fr\'{e} in Turin \cite{Castellani:1991et}. In
this context, the `softening' corresponds to deform the original rigid group
manifold structure of $G_{c}$, where the left- and right- invariant vector
fields and one-forms locally have a fixed coordinate dependence, and where
the Riemannian geometry is (locally) fixed in terms of the structure tensor
of (the Lie algebra $\mathfrak{g}_{c}$ of) $G_{c}$. In this context, it is
worth mentioning that some work of Castellani \cite{ca1,ca2}, in which he
proposed some pioneering generalizations of KM algebras, shares some
intriguing structural properties with the ansatz considered in this paper.
The exact implications deriving from the comparison of both approaches are
not yet fully explored, and their detailed analysis is left for a future
investigation.


Thus, the algebraic generalization achieved in this paper should be regarded
as a relevant structural step towards the application of the generalized KM
algebras in the context of (super)gravity and (super)string theories. As
explicit examples, we have considered the `softening' of the one-sphere $%
\mathbb{S}^{1}$, of the two-sphere $\mathbb{S}^{2}$, and of the three-sphere
$\mathbb{S}^{3}$ (which in particular include the \textit{squashed}
three-sphere $\widetilde{\mathbb{S}}^{3}$, also named \textit{Berger}
three-sphere). While the generalized KM algebra associated to the `softened'
circle $\mathbb{S}_{F}^{1}$ is trivially isomorphic to its undeformed
analogue and thus deprived of interest, the `softening' of $\mathbb{S}^{2}$
and $\mathbb{S}^{3}$ yields to non-trivial results, which would be very
interesting to apply to the broad range of contexts (briefly reviewed within
our treatment) in which their undeformed counterparts play an important role.
In this framework, by developing the suggestion made in the conclusion of
\cite{ca1} (in which a generalization of general relativity for closed
strings was formulated), the generalized KM algebras associated to $\mathbb{S%
}^{2}$ and $\mathbb{S}^{3}$ would be expected to yield an extension of
general relativity for closed 2- and 3- branes, respectively; the physical
meaning of the corresponding `softenings' remains an intriguing venue for
further future research.

Of course, along the lines of research pertaining to the present paper,
 further possible developments would consist into applying the `softening'
procedure to non-compact Lie group manifolds $G_{nc}$ (or cosets thereof) on
which the generalized KM algebras introduced in \cite%
{ram,Campoamor-Strusberg:2024kpl} are based; consequently, $G_{nc}$ would be
`softened' into the manifold $G_{nc}^{\mu }$. Indeed, it should be here
recalled that the prototypical example of `softening' is based on $%
G_{nc}/H=ISO(3,1)/SO(3,1)$, thus realizing the four-dimensional space-time
manifold in which dynamical fields are defined through an
`horizontalization' procedure (cf. e.g. \cite{Castellani:1991et}).
Therefore, the infinite-dimensional algebras resulting from this further
step of algebraic generalization would be based on the non-compact `soft'
group manifold $G_{nc}^{\mu }$, and they would be potentially relevant for
the formulation of gravitational theories, with or without underlying
supersymmetry; in this respect, it would be interesting to investigate
possible relations with the results of the recent paper \cite{Blair:2024ofc}%
. We leave this intriguing venue of investigation for further future
research.

\section*{Acknowledgments}

The work of RCS has been supported by the Agencia Estatal de Investigaci\'on
(Spain) under the grant PID2023-148373NB-I00 funded by
MCIN/AEI/10.13039/501100011033/FEDER, UE. This article is based upon work
from COST Action CaLISTA CA21109 supported by COST (European Cooperation in
Science and Technology).

\appendix

\section{Maurer-Cartan one-forms for $SU(2)$}

\label{sec:App}

For any Lie group (manifold) $G$, the $G$-left-invariance of the `left' $%
\mathfrak{g}$-valued Maurer-Cartan one-form (where $\mathfrak{g}=Lie\left(
G\right) $ is the Lie algebra of $G$) is immediate (see e.g. \cite{Nakahara}%
). Indeed, the `left' Maurer-Cartan one-form $\lambda $ is defined as%
\begin{equation*}
\lambda :=g^{-1}dg,
\end{equation*}%
whereas the left multiplication by an element $h\in G$, denoted as $L_{h} $,
acts on a group element $g\in G$ as $L_{h}\left( g\right) :=hg$. Thus, for
any $g\in G$, the differential of the map $L_{h}$ on the group manifold $G$
is%
\begin{equation*}
d\left( L_{h}\left( g\right) \right) =d\left( hg\right) =hdg.
\end{equation*}%
The invariance of $L_{h}$-transformed one-forms $\lambda ^{\prime }:=\left(
L_{h}\left( g\right) \right) ^{-1}d\left( L_{h}\left( g\right) \right) $
follows at once from the action, as
\begin{equation*}
\lambda ^{\prime }:=\left( L_{h}\left( g\right) \right) ^{-1}d\left(
L_{h}\left( g\right) \right) =\left( hg\right)
^{-1}hdg=g^{-1}h^{-1}hdg=g^{-1}dg=\lambda .~~\square
\end{equation*}

Analogously, one can prove the $G$-right-invariance of the `right' $%
\mathfrak{g}$-valued Maurer-Cartan one-form, defined as%
\begin{equation*}
\omega :=\left( dg\right) g^{-1}.
\end{equation*}%
The right multiplication by an element $h\in G$, denoted as $R_{h}$, acts on
a group element $g\in G$ as $R_{h}\left( g\right) :=gh$. Thus, for any $g\in
G$, the differential of the map $R_{h}$ on the group manifold $G$ is%
\begin{equation*}
d\left( R_{h}\left( g\right) \right) =d\left( gh\right) =\left( dg\right) h.
\end{equation*}%
Again, the $R_{h}$-transformed Maurer-Cartan one-form $\omega ^{\prime
}:=\left( d\left( R_{h}\left( g\right) \right) \right) \left( R_{h}\left(
g\right) \right) ^{-1}$ coincides with itself :%
\begin{equation*}
\omega ^{\prime }:=\left( d\left( R_{h}\left( g\right) \right) \right)
\left( R_{h}\left( g\right) \right) ^{-1}=\left( dg\right) h\left( gh\right)
^{-1}=\left( dg\right) hh^{-1}g^{-1}=\left( dg\right) g^{-1}=\omega
.~~\square
\end{equation*}

The $SU(2) $ group manifold can be parameterised using three \textit{Euler
angles} $\varphi $, $\theta $, and $\psi $, which correspond to a sequence
of rotations about specific axes. As $SU(2) $ is the double cover of the
rotation group $SO(3) $, each rotation in $SO(3) $ corresponds to two points
in $SU(2) $, providing a complete representation of all possible
orientations. The three Euler angles allow us to express any element $g \in
SU(2) $ as a product of rotations as follows:

\begin{equation*}
g(\varphi ,\theta ,\psi )=e^{i\psi \sigma _{3}/2}e^{i\theta \sigma
_{2}/2}e^{i\varphi \sigma _{3}/2}
\end{equation*}%
where $\sigma _{1}$, $\sigma _{2}$, and $\sigma _{3}$ are the Pauli
matrices:
\begin{equation*}
\sigma _{1}=%
\begin{pmatrix}
0 & 1 \\
1 & 0%
\end{pmatrix}%
,\quad \sigma _{2}=%
\begin{pmatrix}
0 & -i \\
i & 0%
\end{pmatrix}%
,\quad \sigma _{3}=%
\begin{pmatrix}
1 & 0 \\
0 & -1%
\end{pmatrix}%
.
\end{equation*}%
The range of the angles are as follows : $\psi \in \lbrack 0,4\pi )$
(initial rotation about the $z$-axis); $\theta \in \lbrack 0,\pi ]$
(rotation about the $y$-axis); $\varphi \in \lbrack 0,2\pi )$ (final
rotation about the $z$-axis). With these parameters, the generic element $%
g(\psi ,\theta ,\varphi )\in SU(2)$ can be written explicitly as:
\begin{equation*}
g(\psi ,\theta ,\varphi )=%
\begin{pmatrix}
e^{i(\psi +\varphi )/2}\cos (\theta /2) & -e^{i(\psi -\varphi )/2}\sin
(\theta /2) \\
e^{-i(\psi -\varphi )/2}\sin (\theta /2) & e^{-i(\psi +\varphi )/2}\cos
(\theta /2)%
\end{pmatrix}%
\in SU(2)\simeq \mathbb{S}^{3},
\end{equation*}%
(to be compared with \eqref{eq:SU2}). This
parameterisation provides a complete description of any element in $SU(2) $
through the three Euler angles $\varphi $, $\theta $, and $\psi $. The
double covering property of $SU(2)$ over $SO(3)$ is reflected in the range
of $\psi $ extending to $4\pi $, ensuring that each rotation in $SO(3)$
corresponds to two points in $SU(2)$.

Using this parameterisation, one can easily compute the right-invariant
Maurer-Cartan one-forms $\omega$'s (\ref{eq:RI1F}). These are obtained by
taking the differential of the group element and transforming it by the
inverse of the group element on the right. Specifically, if $g(\psi ,\theta
,\varphi )$ is a generic element of $SU(2)$, then the right-invariant
Maurer-Cartan one-form $\omega $ is given by:

\begin{equation*}
\omega =(dg)g^{-1}
\end{equation*}%
where $\omega $ is an $\mathfrak{su}(2)$-valued one-form. In this case, we
can expand $\omega $ as a linear combination of the Pauli matrices $\sigma
_{i}$, which form a basis for $\mathfrak{su}(2)=Lie\left( SU(2)\right) $.
Thus, one obtains

\begin{equation*}
\omega =\frac{i}{2}\left( \omega _{1}\sigma _{1}+\omega _{2}\sigma
_{2}+\omega _{3}\sigma _{3}\right)
\end{equation*}%
where $\omega _{1}$, $\omega _{2}$, and $\omega _{3}$ are given by (\ref%
{eq:RI1F}). Note that $\omega _{3}$ is also invariant under $L_{0}$.

Analogously, one can compute the left-invariant Maurer-Cartan one-forms $%
\lambda$'s (\ref{eq:LI1F}), defined as%
\begin{equation*}
\lambda =g^{-1}dg=\frac{i}{2}\left( \lambda _{1}\sigma _{1}+\lambda
_{2}\sigma _{2}+\lambda _{3}\sigma _{3}\right) .
\end{equation*}%
Note that $\lambda _{3}$ is also invariant under $R_{0}$. An equivalent way
to obtain the one-forms $\lambda $'s is to observe that%
\begin{equation*}
\left. R_{\pm }\right\vert _{\varphi \leftrightarrow \psi }=-L_{\pm
},~\left. R_{0}\right\vert _{\varphi \leftrightarrow \psi }=L_{0},
\end{equation*}%
and thus ($\forall i=1,2,3$)
\begin{equation*}
\lambda _{i}=\left. \omega _{i}\right\vert _{\varphi \leftrightarrow \psi },
\end{equation*}%
as it follows immediately from (\ref{eq:RI1F}) and (\ref{eq:LI1F}).

\bibliographystyle{utphys}
\bibliography{ref}

\end{document}